\documentclass[aps,pra,reprint,superscriptaddress,noeprint]{revtex4-2}

\usepackage[english]{babel}
\usepackage{tabularray}
\usepackage{tabularx} 
\usepackage{multirow}
\usepackage{amsmath}
\usepackage{amssymb}
\usepackage{wasysym}
\usepackage{bm}
\usepackage{graphicx}
\usepackage{hyperref}
\usepackage{xcolor}
\usepackage{csquotes}
\usepackage{xfrac}
\usepackage{physics}
\usepackage{units}
\usepackage[normalem]{ulem}
\usepackage[capitalise]{cleveref}

\begin{document}

\title{Dissipationless tune-out trapping for a lanthanide-alkali quantum gas mixture}

\author{Alexandre De Martino}
\email[]{These authors contributed equally to this work.}
\affiliation{Physikalisches Institut, Eberhard Karls Universit\"at T\"ubingen, 72076 Tübingen, Germany}

\author{Florian Kiesel}
\email[]{These authors contributed equally to this work.}
\affiliation{Physikalisches Institut, Eberhard Karls Universit\"at T\"ubingen, 72076 Tübingen, Germany}

\author{Jonas Auch}
\affiliation{Physikalisches Institut, Eberhard Karls Universit\"at T\"ubingen, 72076 Tübingen, Germany}

\author{Kirill Karpov}
\affiliation{Physikalisches Institut, Eberhard Karls Universit\"at T\"ubingen, 72076 Tübingen, Germany}

\author{Christian Gross}
\email[]{christian.gross@uni-tuebingen.de}
\affiliation{Physikalisches Institut and Center for Integrated Quantum Science and Technology, Eberhard Karls Universit\"at T\"ubingen, 72076 Tübingen, Germany}

\date{\today}

\begin{abstract}
	Quantum gas mixtures offer a wide field of research, ranging from few-body physics of impurities to many-body physics with emergent long-range interactions and ultracold molecular gases.
	Achieving precision control of mixtures is much harder than for single-component gases and, consequently, the respective techniques are less developed.
	Here we report on a decisive step forward in this direction by realizing dissipationless and fully differential optical control of the motional degrees of freedom of one of the species without affecting the other.
	This is achieved with $^{166}\mathrm{Er}$ and $^6\mathrm{Li}$ in a novel dual-species experiment, where the measurements are performed on each species separately.
	Our experiments pave the way to a new generation of precision many-body experiments with quantum gas mixtures with unprecedented long lifetimes and low temperatures.
\end{abstract}

\maketitle

Ultracold quantum gas mixtures provide a unique platform to explore heterogeneous quantum many-body systems~\cite{baroni2024}.
They have been successfully used to study ultracold molecular \cite{cornish2024, langen2024}, polaronic~\cite{scazza2022, grusdt_25}  and few-body \cite{naidon2017} physics or to realize many-body systems with new forms of emergent interactions \cite{desalvo2019, baroni2024a, cai_2025}.
A ubiquitous difficulty for precision experiments with such mixtures is to realize dissipationless independent control of the motional degree of freedom of the involved atomic species.
In principle, independent control can be achieved with lasers at particular species-dependent wavelengths, so called tune-out wavelengths.
At these wavelengths, the optical polarizability of one of the species vanishes~\cite{leblanc07, catani09}.
Tune-out wavelengths have been measured and used in many experiments with various atomic species~\cite{copenhaver19, decamps20, holmgren12,mckay2013, trubko17, arora11, herold12, leonard15, schmidt16, rubioabadal19, ratkata21, henson15, heinz20a, hoehn23, hoehn24, kao17}, and heteronuclear mixtures~\cite{catani09,vaidya2015, hewitt2024, lippi24}.
Specifically for mixtures, all systems realized so far suffer from a central issue:
the tune-out wavelengths emerge due to destructive interference of the polarizability contribution of atomic transitions, which are not far detuned enough to prevent excess heating of at least one of the species due to photon scattering.
In a tune-out trapped mixture, the relevant parameter to measure the impact of dissipation is $\gamma/U$ the photon scattering rate of both partners $\gamma$ normalized to the potential experienced by the trapped species $U$.
In previously reported experiments the tune-out was realized in alkalis, either in-between the D-lines with Rb-K~\cite{catani09} and Cs-Li~\cite{lippi24}, or using the blue lines to cancel the polarizability in Rb-Yb~\cite{vaidya2015}.
In all these mixtures, $\gamma/U$ is at least one order of magnitude larger than usually achieved in single species experiments.
In contrast, the lanthanide elements feature very narrow and closed transitions lines far in the infrared, ideal to realize dissipationless differential control when combined with alkali atoms.

\begin{figure}[t]
	\centering
	\includegraphics[scale=1]{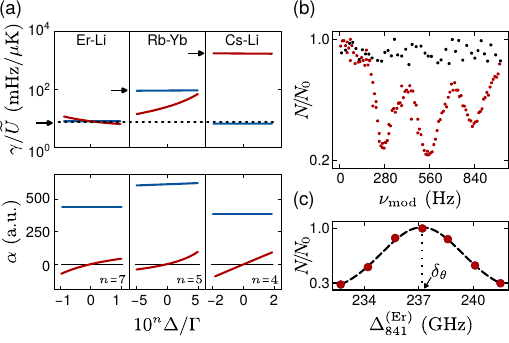}
	\caption{
		System comparison and tune-out spectroscopy.
		(a) We compare three mixtures, for which tune-out trapping has been realized regarding their normalized photon scattering $\gamma/\tilde{U}$ (top) and their polarizabilities $\alpha$ in atomic units (a.u.) (bottom).
		The normalization by $\widetilde U=U/k_\text{B}$, with $k_\text{B}$ being Boltzmann's constant, is done with regard to the potential $U$ seen by the trapped species.
		The arrows in the top panel indicate the relevant value dominated by the most dissipative partner, and the dashed line is a reference value for the often realized single species configuration: Rb trapped in $\unit[1064]{nm}$ light.
		In both panels, the red lines indicate the tuned-out element (Er, Rb, Cs), the blue lines the trapped partners (Li, Yb, Li).
		The horizontal axis displays the normalized and scaled detuning $10^n \Delta/\Gamma$, where $\Delta$ is the frequency difference to the tune-out point (for Er: $\unit[841]{nm}$, Rb: $\unit[423]{nm}$, Cs: $\unit[880]{nm}$), $\Gamma$ is the line width of the nearest transition line of the tuned-out element and $n$ accounts for the different distance of this line to the tune-out point.
		We use the polarizability of Li and Cs from~\cite{UDportal}.
		(b) Trap-loss spectrum of Er in the $\unit[1064]{mm}$ optical trap beam, which is superimposed with a $\unit[841]{nm}$ dimple beam.
		The intensity of the dimple beam is amplitude-modulated at different frequencies $\nu_\text{mod}$.
		The red data is representative for a non-fine-tuned dimple wavelength away from the tune-out, the black data is on the tune-out.
		We attribute the slight modulation of the black data to experimental instabilities during the measurement.
		(c) Remaining normalized Er atom number (red dots) for resonant modulation at the frequency of the first minimum in (b) for different dimple beam detunings from the $\unit[841]{nm}$ line.
		The tune-out shown here at $\unit[237]{GHz}$ is close to the maximum detuning of $\unit[245]{GHz}$ for $\theta=90^\circ$.
		Error bars of one standard deviation are smaller than the marker size.
		A Gaussian fit (dashed line) is used to extract the tune-out wavelength $\delta_\theta$ and its uncertainty.
	}
	\label{fig:fig1}
\end{figure}

Here we report on the measurement of a tune-out wavelength for $^{166}$Er near its $\unit[841]{nm}$-line (see \cref{fig:fig1}) which we combine with $^{6}$Li in our dual-species experiment.
We observe very low dissipation for both Er and Li, as expected for the large effective detuning of the tune-out wavelength to all transitions.
We precisely determine the dependence of the tune-out wavelength on the relative angle between polarization and magnetic field.
This provides accurate spectroscopic data for the tensor light shift in Er, which significantly deviates from the predictions based on the line data published in~\cite{becher18}.
Our experiments pave the way for a new generation of dual species precision experiments with fully differential control over the motional degrees of freedom.
This allows for the experimental study of motional properties of impurities with extreme mass imbalance~\cite{christianen2022b}, the implementation of novel sympathetic cooling schemes to achieve lower reduced temperatures for ultracold fermions~\cite{onofrio02, presilla03}, and the study of fermionic lattice systems in contact with a bosonic bath~\cite{griessner2006, bruderer2008a}.

Our experiment features spatially separated magneto-optical traps for Er and Li, which are placed in line of sight with each other and a glass cell providing high optical access.
Both gases are transported into this cell in a running optical lattice realized with $\unit[1064]{nm}$ laser light.
Li is transported over a distance of $\unit[1]{m}$, Er over $\unit[0.5]{m}$ using two Gaussian beams with displaced foci~\cite{matthies2024}.
Additionally, Er is magnetically levitated against gravity during transport.
In all measurements reported here, we either work with Er or with Li, loaded into orthogonally crossed single-mode laser beams with beam waists of $\unit[35]{\mu m} \times \unit[580]{\mu m}$ at $\unit[1064]{nm}$ after transport.
This provides a trap with a depth of $\unit[36]{\mu K}$ for Li and $\unit[15]{\mu K}$ for Er.
The large aspect ratio is required to prevent heating during loading from the lattice and to confine Er against gravity.
A vertical magnetic field of a few Gauss is used to conserve the spin alignment of the gases. For a more detailed experimental description, see the Supplemental Material \cite{sm} (see also references \cite{aikawa12, phelps20, lunden20, plotkinswing20} therein).
We superimpose a smaller dimple laser beam at $\unit[841]{nm}$ derived from a titanium-sapphire laser with waists of $\unit[6]{\mu m}\times\unit[28]{\mu m}$ and with a power of up to $\unit[0.5]{W}$.
This beam propagates in a direction perpendicular to the magnetic field, and its polarization is aligned to be linear with an angle $\theta=90^\circ$ to it.
The laser is frequency stabilized to a low-drift reference cavity and tunable over a range of several $\unit[100]{GHz}$ with negligible frequency-error in this range.

The size of the dimple is chosen such that for the same atom numbers, a degenerate Fermi gas of Li trapped in it is fully immersed in a Bose-Einstein condensate of Er held in the large $\unit[1064]{nm}$ trap.
The separation of scales, even more in single particle eigenenergies than in trap size, is advantageous: the dimple trap realizes a tune-out for Er only, but Li, being much more tightly trapped, is essentially unaffected by the small gradients of the large trap.
A separation of scales is present in many properties of a Er-Li mixture.
In contrast to Li, Er is very heavy and magnetic.
Li features few very broad Feshbach resonances~\cite{zurn2013a}, while Er has many much narrower ones~\cite{frisch2014} and also first studies of interspecies resonances have been reported~\cite{schafer2022}.
This is an ideal starting point to explore optimized sympathetic cooling of the fermions with a largely populated coolant, that can initially be prepared below one-hundreds of the fermi temperature~\cite{presilla2003}.
Recent experiments with a similarly mass-imbalance Dy-Li mixture~\cite{xie2025} and a dual Bose gas of Er-Li~\cite{kalia2025} have already demonstrated thermalization between the two components.

To measure the tune-out wavelength, we modulate the power of the dimple beam with 100\% amplitude-variation and variable frequency (see~\cref{fig:fig1}).
After about $\unit[1]{s}$ of modulation we measure the remaining atom number.
Without fine-tuning of the dimple wavelength, we observe three resonances corresponding to the trap frequency in the $\unit[1064]{nm}$ dipole trap at $\unit[280]{Hz}$ and its first two harmonics at $\unit[460]{Hz}$ and $\unit[840]{Hz}$.
In contrast, no effect of the modulation is visible when the dimple trap wavelength is set to the tune-out value.
To precisely locate it relative to the $\Gamma^\text{(Er)}_\text{841} = 2\pi \times \unit[8]{kHz}$ wide transition at $\unit[841.22]{nm}$, we measure the remaining atom number versus the detuning $\Delta^\text{(Er)}_\text{841}$.
A fit to the data reveals the tune-out point for this geometry at
$\delta_{90^\circ}=\unit[245 \pm 1]{GHz}$ and we experimentally confirmed that the results are independent of the intensity of the $\unit[1064]{nm}$ trap.
The found tune-out frequency is far blue detuned $\delta_{90^\circ} \approx \unit[3.1\cdot10^7]{\Gamma^\text{(Er)}_\text{841}}$ to the $\unit[841]{nm}$-line of Er.

\begin{figure}[t]
	\centering
	\includegraphics[scale=1]{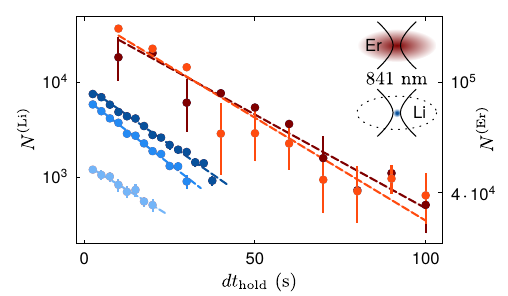}
	\caption{
		Dissipative effects of the tune-out beam on Er and Li.
		Lifetime measurements of Er in the $\unit[1064]{nm}$ optical dipole trap are shown for the $\unit[841]{nm}$ dimple beam set to the tune-out wavelength at full (dark red) and zero power (light red).
		The right vertical axis gives the trapped Erbium  atom number $N^\text{(Er)}$ versus holding time $dt_{\text{hold}}$.
		Each data point is averaged over five experimental repetitions.
		Lifetimes are extracted by an exponential fit (dashed lines) to the data.
		In blue, we show lifetime measurements of Li in the same dimple beam for three powers: $\unit[(0.1, 0.3, 0.5)]{W}$ (light to dark blue).
		In-trap atom numbers for Li $N^\text{(Li)}$ are given on the left vertical axis, and each point is the average of ten experimental runs.
		The extracted lifetimes are found to be power-independent within the experimental uncertainties.
		For all measurements, the error bars represent the standard error of the mean.
		The two different measurement settings are sketched in the upper right corner inset.
	}
	\label{fig:lifetimes}
\end{figure}

To quantify the dissipative effect of the laser light at the tune-out wavelength due to photon scattering, we measure the lifetime of Er in the $\unit[1064]{nm}$ trap while exposed to the dimple light.
We set the dimple power to $\unit[0.5]{W}$, which corresponds to a trap depth for Li of $U^\text{(Li)} = k_{\text{B}} \cdot \unit[0.2]{mK}$.
The dimple light is linearly polarized, and the polarization axis is set at the angle $\theta=90^\circ$ with respect to the quantization axis.
The measured lifetimes, shown in \cref{fig:lifetimes}, demonstrate the non-dissipative character of our dimple trap for Er.
For holding times up to $\unit[100]{s}$, we do not observe any detectable change of the 1/e-lifetime.
We measure $\tau^\text{(Er)}_\text{841} = \unit[66 \pm 16]{s}$ with and $\tau^\text{(Er)} = \unit[58 \pm 18]{s}$ without dimple.
Our observation matches the theoretical prediction:
photon scattering with rate $\gamma$ transmits energy to the trapped gas according to $\dot{E} = 2 E_r \gamma$, where $E_r = (\hbar k)^2/2m$ is the recoil energy imparted on the atoms of mass $m$ from the scattered photons with wave number $k$.
The photon scattering rate for a laser of angular frequency $\omega_L$ follows as the sum over the contributions from all far-detuned transition lines $\gamma=\sum_i \Gamma_i \Omega_i^2/4\Delta_i^2$ with Rabi frequency $\Omega_i$, detuning $\Delta_i=\omega_L-\omega_{i}$ and inverse lifetime $\Gamma_i$ of transition line $i$.
We overestimate the scattering rate by assuming a homogeneous dimple peak intensity over the entire Er cloud and obtain a scattering rate of $\gamma^\text{(Er)}=\unit[1.5]{Hz}$, corresponding to an energy increase rate of $\dot{E}^\text{(Er)}=k_B \cdot \unit[250]{n K/s}$.
Normalizing these numbers to the dimple trap depth for Li, we obtain $\gamma^\text{(Er)}/U^\text{(Li)}=\unit[8]{mHz/\mu K} \, k_\text{B}$ or ${\dot{E}^\text{(Er)}/U^\text{(Li)}= \unit[1.3]{n K/\mu K \, s}}$.
The geometric overlap of the Er cloud and dimple beam is only $6\%$, leading to an effective heating rate of $\dot{E}^\text{(Er)}_{\text{eff.}}=k_B \cdot \unit[15]{n K/s}$.
These calculations are based on the Er line data published in~\cite{becher18} with the theory data replaced by the experimental values for the lines at $\unit[401]{nm}$, $\unit[583]{nm}$ and $\unit[841]{nm}$~\cite{frisch14, hartog10,ban05}.

We also measure the dissipative effects of the dimple on Li trapped in it.
To this end, we prepare Li in a balanced $\ket{F,m_F}=\ket{1/2,1/2}$, $\ket{3/2,-3/2}$ hyperfine state mixture and measure its lifetime for different dimple trap powers (see \cref{fig:lifetimes}).
We observe a power-independent 1/e-lifetime with a mean value of $\tau_{841}^{\text{(Li)}} = \unit[15.7 \pm 1.9]{s}$.
The power independence (also observed for Er and Li in the $\unit[1064]{nm}$ trap) suggests that we reach the fundamental limit given by photon scattering, for which the imparted energy and trap depth both scale linearly with the laser intensity.
The off-resonant scattering for Li is dominated by its $\unit[671]{nm}$ line, which is $\Delta^\text{(Li)}_\text{671} \approx 1.5 \cdot 10^7 \,\Gamma^\text{(Li)}_\text{671}$ detuned.
The photon scattering rate is the same as for Er: $\gamma^\text{(Li)}/U^\text{(Li)} = \unit[8]{mHz/\mu K} \, k_\text{B}$, but the energy input rate is 28-fold increased to $\dot{E}^\text{(Li)}/U^\text{(Li)}= \unit[40]{n K/\mu K \,s}$ due to the smaller mass.
It is important to note that the energy input rate is not always the relevant number.
For small sample sizes of about 100 atoms, the timescale provided by $\gamma$ can be long enough to perform experiments without any photon scattering.

\begin{figure}[t]
	\centering
	\includegraphics[scale=1]{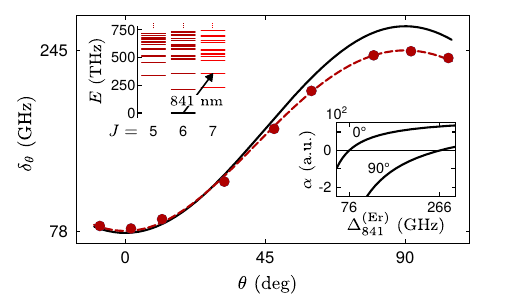}
	\caption{
		Polarization dependence of the tune-out wavelength.
		Measured tune-out frequencies (red dots) are given as detuning from the $\unit[841]{nm}$-line for different angles between the linear polarization $\theta$ and the magnetic field.
		The tune-out frequencies $\delta_\theta$ change sinusoidally in dependence of $\theta$ in an interval of $\unit[165]{GHz}$.
		The contributing lines up to $\unit[400]{nm}$ are visualized in the energy level diagram (inset, top left), where the lines connecting to the dark red states cause the $\theta$-dependence.
		The inset on the bottom right displays the calculated polarizability for parallel and perpendicular polarized linear light, visualizing the shift of the zero-crossing/tune-out point for the two extreme configurations.
		While the theory (black line) matches the measurements for small angles, a significant deviation becomes apparent closer to perpendicular alignment of polarization and magnetic field.
		We use a fit (dashed red line) to quantify the origin of this deviation, as detailed in the main text.
	}
	\label{fig:polarization_angle}
\end{figure}

The tune-out wavelength is expected to show a strong dependence on the polarization angle $\theta$ between the linearly polarized light and the magnetic field~\cite{lepers14, becher18}, which has been studied in Dy~\cite{kao17}.
This is due to the tensor polarizability of the lanthanides and, in Er, due to the sizable contribution to the tensor polarizability from lines connecting to states with electronic angular momentum differing from $J'=7$, the value for the $\unit[841]{nm}$-line.
The most important ones are the broad $J'=5,6$ lines in the blue part of the spectrum, and with $\theta$ the coupling strength to these lines varies.
To measure the polarization dependence, we change $\theta$ and repeat the measurement process described above (see \cref{fig:polarization_angle}).
We concentrate on linear polarization,
eliminating the effect of the vector light shift~\cite{le_kien_2013, li17} and observe a sinusoidal shift of the tune-out frequency between $\delta_{0^\circ} = \unit[80]{GHz}$ and $\delta_{90^\circ}=\unit[245]{GHz}$.
A comparison of our measurements with the theoretical prediction, based on the same modified line data set as above, reveals a discrepancy in the amplitude of the $\theta$ dependent modulation.
We note that the unmodified line data of~\cite{becher18} cannot be matched with our measurements.
To quantify the theory-experiment mismatch, we fix the contributions of the well characterized $\unit[841]{nm}$-line to the prediction of the scalar ($\alpha_{s,\text{841}}$) and tensor ($\alpha_{t,\text{841}}$) polarizability.
We then decompose the polarizabilities into contributions from this close-by line and all other far-detuned lines by writing $\alpha_s =  \alpha_{s,\text{841}} / \Delta + \alpha_{s,0}$ and $\alpha_t = \alpha_{t,\text{841}} / \Delta  + \alpha_{t,0}$ with the far-detuned \enquote{background} contributions being constant over the explored frequency interval.
We formulate the tune-out condition as $\delta_\theta = - (\alpha_{s,\text{841}} + \tfrac{1}{2} [3 \cos^2(\theta)-1] \alpha_{t,\text{841}} )/(\alpha_{s, 0} + \tfrac{1}{2} [3 \cos^2(\theta)-1] \alpha_{t,0})$.
Fitting this equation to our measurement, results in $\alpha_{s,0}= \unit[(193 \pm 5)]{a.u.} $ and $\alpha_{t,0}= \unit[(-17 \pm 0.4)]{a.u.}$, with $\unit[1]{a.u.} = 4\pi \epsilon_0 a_b^3$ expressed in the Bohr radius $a_b$ and the vacuum electric permittivity $\epsilon_0$.
The error is dominated by the uncertainty of the line width $\Gamma^\text{(Er)}_\text{841}=\unit[(8.0 \pm 0.2)]{kHz}$ \cite{ban05}.
Comparing the fitted background polarizabilities to the theoretical prediction $\alpha_{s,0}^{th}=\unit[184]{a.u.}$ reveals little relative difference for the scalar part, but a 10-fold difference in the tensor part $\alpha_{t,0}^{th}= \unit[-1.6]{a.u.}$

We have demonstrated a dissipationless tune-out wavelength for Er, which at the same time allows far-detuned conservative trapping of Li.
Our results demonstrate a reduction of the dissipative effects of a tune-out wavelength by one order of magnitude compared to previously realized quantum gas mixtures~\cite{vaidya2015}.
This is an important milestone en route to precision control of heteronuclear quantum gas mixtures at the coldest possible temperatures.
The combination of Er with Li is very flexible for the future.
It allows realizing heavy-light boson-fermion, boson-boson and fermion-boson systems with tunable interactions, of which the Bose-Bose mixture recently has been brought to degeneracy~\cite{kalia2025}.
Er features an even narrower transition line at $\unit[1299]{nm}$~\cite{patscheider_2021}, to which our idea can also be applied.
This would result in a further reduction of the dissipation by more than one order of magnitude.

\textit{Acknowledgments}\\
We acknowledge funding from the Horizon Europe program HORIZON-CL4-2022-QUANTUM-02-SGA via the project 101113690 (PASQuanS2.1), the Federal Ministry of Education and Research Germany (BMBF) via the project FermiQP (13N15889), the Deutsche Forschungsgemeinschaft within the research unit FOR5522 (Grant No. 499180199) and the Alfried Krupp von Bohlen and Halbach foundation.

\textit{Data availability}\\
The experimental and theoretical data and evaluation scripts that support the findings of this study are available on Zenodo~\cite{zenodo}.

\newpage
\clearpage

\section*{Supplemental Material}

\subsection*{Experimental sequence}

For the characterization of the tune-out for Er, our experiments start with the preparation of the trapped Er cloud.
The procedure follows the standard approach for this atom~\cite{aikawa12, phelps20}.
We use a spin-flip Zeeman slower on the strong line of Er at $\unit[401]{nm}$ ($\Gamma=2\pi\cdot \unit[29.7]{MHz} $), which is directly loading a narrow-line magneto-optical trap (MOT) at $\unit[583]{nm}$ ($\Gamma=2\pi\cdot \unit[190]{kHz}$).
An important addition to this setup was the use of \enquote{angled slowing} beams~\cite{lunden20, plotkinswing20}.
Two beams, red-detuned from the $\unit[401]{nm}$-transition, were aligned to cross just before the MOT region and provide an intermediate slowing step, after the slower's exit and right in front of the MOT's capture region.
In our setup, this additional cooling improved the MOT loading rate by more than an order of magnitude, reaching $4\cdot10^{7}\mathrm{/s}$.
After a loading time of $\unit[5]{s}$, the compressed MOT (cMOT) contains a approximately $2\cdot10^{8}$ atoms, cooled down to $\unit[8]{\mu K}$ and optically pumped into the lowest Zeeman state $\ket{J, m_J}=\ket{6, -6}$.

The atoms are then transferred from the cMOT into a crossed-beam optical dipole trap (ODT), consisting of two laser beams at $\unit[1064]{nm}$, focused to a waist of $\unit[80]{\mu m}$ at $\unit[30]{W}$ each.
To this trap, an additional beam used later for transport is superimposed, also at $\unit[1064]{nm}$ and with a e$^{-2}$ radius of $\unit[500]{\mu m}$ at $\unit[80]{W}$.
To transport this cloud into the glass cell, we ramp up a second transport beam to $\unit[30]{W}$ to create a horizontal lattice.
To transport this cloud into the glass cell, we ramp up a second transport beam to $\unit[30]{W}$ to create a horizontal lattice.
We then ramp down the ODT power and move the lattice over $\unit[0.5]{m}$ in $\unit[165]{ms}$ by detuning one of the lattice beams.
However, this lattice is not strong enough to hold the Er atoms against gravity, and we use two elongated coils (\enquote{racetrack coils}) to generate a vertical magnetic field gradient of $\unit[4.2]{G/cm}$ at a vertical offset field of $\unit[22]{G}$.
This levitates the cloud over the entire transport distance.
Finally, the cloud is transferred in an elliptical ODT consisting of two laser beams of vertical and horizontal waists of $(\omega_v, \omega_h)=\unit[(35, 580)]{\mu m}$, at $\unit[1064]{nm}$ and $\unit[10]{W}$ each.
The vertical confinement is strong enough to switch off the levitation field, while the vertical offset field is reduced to $\unit[4.9]{G}$.

The production of an ultra-cold gas of Li and its transport to the glass cell follows similar steps as for Er.
We start with a MOT loaded from a decreasing-field Zeeman slower, both working on the $\mathrm{D_2}$-line at $\unit[671]{nm}$ ($\Gamma=2\pi\cdot \unit[5.9]{MHz}$~\cite{McAlesander96}).
An important, non-standard feature of our setup is the implementation of a transverse cooling and optical pumping stage before the Zeeman slower input.
By using a two-dimensional molasses on the $\mathrm{D_1}$-line, the atoms are pumped into the hyperfine $\ket{F, m_F}=\ket{3/2, -3/2}$ state, which is the only one being addressed by the Zeeman slower.
The combined contributions of pumping and transversal cooling result in a ten-fold increase in the MOT loading rate, yielding a value of $2\cdot10^{7}\mathrm{/s}$.
After a loading time of $\unit[5]{s}$, and at the end of a cMOT stage, we measure approximately $10^{8}$ atoms, cooled down to $300\mu\mathrm{K}$.

The Li cloud is then transferred to a $\unit[1070]{nm}$ multimode ODT consisting of two horizontal beams overlapped at an angle of $10^\circ$, with powers and waists of $\unit[150]{W}$ and $\unit[65]{\mu m}$.
The Li cloud is in an equal spin mixture of $\ket{F, m_F} = \ket{1/2, \pm1/2}$.
It is further cooled by evaporating at $\unit[320]{G}$ by ramping down the laser power for $\unit[2.4]{s}$.
At this stage, we reach degeneracy when evaporating further.
Right before the end of evaporation, we ramp up the transport lattice, and transport the atoms over a distance of $\unit[1]{m}$ into the glass cell.
The transport of Li takes $\unit[112]{ms}$.
When the atoms have arrived in the glass cell, we transfer the population from state $\ket{1/2, -1/2}$ to state $\ket{3/2, -3/2}$.
This is being done by a rapid adiabatic passage with a $\unit[100]{kHz}$ wide sweep of a rf-field over the resonance at $\unit[3.4]{G}$ in $\unit[25]{ms}$.
Finally, the cloud is transferred to the same elliptical trap as Er before, and then loaded into the tune-out dimple trap at $\unit[841]{nm}$ with waists of $(\omega_v, \omega_h)= \unit[(6, 28)]{\mu m}$ .

\begin{figure*}[t]
	\includegraphics[scale=1]{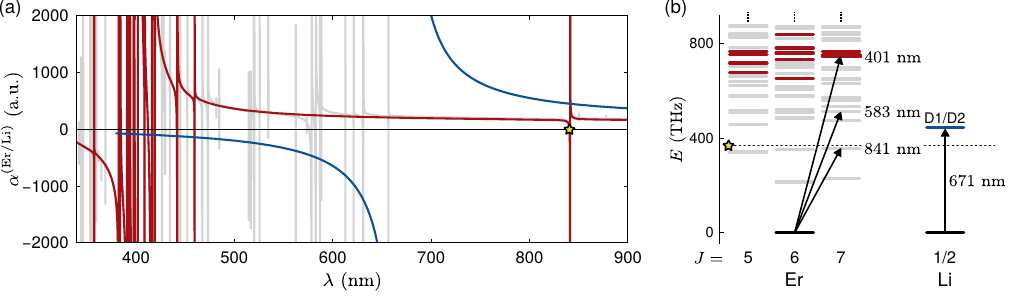}
	\caption{
		(a) Polarizabilities of Er and Li, with the relevant tune-out highlighted (star).
		The total polarizability of Er (gray) consists of contributions from many lines, but it is dominated by strong lines in the blue and UV.
		Those lines together with the $\unit[841]{nm}$ line (red) govern the position of the tune-out wavelength, which is blue detuned from this $2\pi \times \unit[8]{kHz}$ narrow line.
		The polarizability of Li (blue) is dominated by the $\unit[671]{nm}$ $\mathrm{D_1}$ and $\mathrm{D_2}$ lines.
		The tune-out point near $\unit[841]{nm}$ of Er is far red-detuned to them, which allows creating attractive optical dipole traps for Li at that wavelength.
		(b) Energy levels for Er and Li. Arrows indicate commonly used cooling transitions.
		Strong and weak Er lines are colored the same way as in (a).
	}
	\label{fig:polarizabilities}
\end{figure*}

\subsection*{Theory of polarizability / tune-out}
When exposed to a light field, an atom gets polarized and its electronic cloud's distribution is modified.
The atom's response generally depends on its state and dipole-allowed transitions, but also on the wavelength, polarization, and relative orientation of the light to the atom.
This interaction results in an energy shift of the atom proportional to the intensity $I=\frac{\epsilon_0 c}{2}|\bm{E}|^2$ of the light field.
Here $\bm{E}$ is the electric field, $\epsilon_0$ the vacuum permittivity and $c$ the speed of light.
Suppose linear polarized light, the angle between the polarization and the quantization axis (usually given by an offset magnetic field) is denoted by the angle $\theta$ and the angle between the propagation of the light field and the quantization axis is denoted by $\kappa$.
The total polarizability $\alpha_{tot}$ of an atom and its energy shift can be decomposed into three contributions \cite{li17} of distinct character, respectively:
\begin{equation} \label{eq:u_tot}
	\begin{split}
		U(\omega) = & -\frac{1}{2 \epsilon_0 c}I(r) \alpha_{tot}                                                                                                                            \\
		=           & -\frac{1}{2 \epsilon_0 c}I(r) \biggr[ \alpha_s(\omega) + |\bm{u}^* \times \bm{u}| \cos(\kappa) \frac{m_J}{2J}  \alpha_v(\omega)                                       \\
		            & + \underbrace{\frac{3m_J^2-J(J+1)}{J(2J-1)}}_\text{= 1, $\forall \space \space |m_J| = J \neq 1/2$   }  \frac{3 \cos^2(\theta)-1}{2}\alpha_t(\omega) \biggr] \text{.}
	\end{split}
\end{equation}
The scalar contribution $\alpha_s (\omega)$ depends solely on the wavelength of the light, and is the main contributor to the polarizability of alkali atoms.
It is the same for all Zeeman substates.
The vector contribution $\alpha_v (\omega)$ acts as a fictitious magnetic field.
Its impact vanishes for linear polarized fields, as the cross product $|\bm{u}^* \times \bm{u}| = 0$ of a real polarization vector.\newline
The tensor contribution $\alpha_t (\omega)$ originates from an induced polarization perpendicular to the electric field wave vector of the driving field.
It does not contribute to atoms in $J=1/2$ states, which holds for all alkali ground states.
It also vanishes for $\theta_{0} = 54.7^\circ$.
For alkaline-earth and lanthanide atoms, a more complex electron configuration leads to high ground state total angular moment numbers of, e.g., $J=6$ for Er, and thus generally to a non-vanishing contribution of $\alpha_t$.
The three individual contributions are given by:
\begin{equation} \label{eq:alphas}
	\begin{split}
		\alpha_s(\omega) & =-\sqrt{\frac{1}{3(2J+1)}} \alpha_J^{(0)}(\omega) \text{,}                   \\
		\alpha_v(\omega) & =+\sqrt{\frac{2J}{(J+1)(2J+1)}} \alpha_J^{(1)}(\omega) \text{,}              \\
		\alpha_t(\omega) & =+\sqrt{\frac{2J(2J-1)}{3(J+1)(2J+1)(2J+3)}} \alpha_J^{(2)}(\omega) \text{,}
	\end{split}
\end{equation}
with
\begin{equation} \label{eq:alpha_k}
	\begin{split}
		\alpha_J^{(K)}(\omega) = & \sqrt{2K+1} \times \sum\limits_{J^\prime}^{}(-1)^{J+J^\prime}                                     \\
		                         & \times\begin{Bmatrix}
			                                 1 & K          & 1 \\
			                                 J & J^{\prime} & J \\
		                                 \end{Bmatrix}
		|\bra{J^{\prime}}|\textbf{d}|\ket{J}|^2                                                                                      \\
		                         & \times \frac{1}{\hbar}\text{Re} \left[ \frac{1}{\Delta^{-}_{J^\prime J} - i\gamma_{J^\prime}/2} +
			\frac{(-1)^{K}}{\Delta^{+}_{J^\prime J} - i\gamma_{J^\prime}/2}\right] \text{.}
	\end{split}
\end{equation}
The sum includes all dipole allowed transitions from $J$ to $J^\prime$ ($\Delta J= \pm 1,0$).
The curly brackets matrix denotes the Wigner 6-j symbol.
We calculate the reduced dipole transition element $|\bra{J^{\prime}}|\bm{d}|\ket{J}|$ as in \cite{le_kien_2013}.
$\gamma_{J^\prime}$ is the excited state natural line width and $\Delta^{\pm}_{J^\prime J}= \omega_{J^\prime J} \pm \omega$.

Our calculations are based on the theoretical line data provided in \cite{becher18}, which we modified with experimentally measured values for the lines $\unit[841]{nm}$ \cite{ban05}, $\unit[583]{nm}$\cite{hartog10} and $\unit[401]{nm}$ \cite{frisch14}.
In the following, we assume that the light travels perpendicular to the magnetic field ($\kappa=90^\circ$), which leaves us with only the scalar and tensor contributions of \cref{eq:u_tot}.
We take the polarizability of Li from \cite{UDportal} and display the Er and Li polarizabilities together in \cref{fig:polarizabilities}.
The isolation of the $\unit[841]{nm}$ line from the strong lines in the blue is apparent.
On the tune-out point of Er, the Li $\unit[671]{nm}$ lines are far red-detuned leading to an attractive optical potential.

\newpage

\bibliography{references.bib}

\providecommand{\noopsort}[1]{}\providecommand{\singleletter}[1]{#1}%
\begin{thebibliography}{58}%
\makeatletter
\providecommand \@ifxundefined [1]{%
 \@ifx{#1\undefined}
}%
\providecommand \@ifnum [1]{%
 \ifnum #1\expandafter \@firstoftwo
 \else \expandafter \@secondoftwo
 \fi
}%
\providecommand \@ifx [1]{%
 \ifx #1\expandafter \@firstoftwo
 \else \expandafter \@secondoftwo
 \fi
}%
\providecommand \natexlab [1]{#1}%
\providecommand \enquote  [1]{``#1''}%
\providecommand \bibnamefont  [1]{#1}%
\providecommand \bibfnamefont [1]{#1}%
\providecommand \citenamefont [1]{#1}%
\providecommand \href@noop [0]{\@secondoftwo}%
\providecommand \href [0]{\begingroup \@sanitize@url \@href}%
\providecommand \@href[1]{\@@startlink{#1}\@@href}%
\providecommand \@@href[1]{\endgroup#1\@@endlink}%
\providecommand \@sanitize@url [0]{\catcode `\\12\catcode `\$12\catcode `\&12\catcode `\#12\catcode `\^12\catcode `\_12\catcode `\%12\relax}%
\providecommand \@@startlink[1]{}%
\providecommand \@@endlink[0]{}%
\providecommand \url  [0]{\begingroup\@sanitize@url \@url }%
\providecommand \@url [1]{\endgroup\@href {#1}{\urlprefix }}%
\providecommand \urlprefix  [0]{URL }%
\providecommand \Eprint [0]{\href }%
\providecommand \doibase [0]{https://doi.org/}%
\providecommand \selectlanguage [0]{\@gobble}%
\providecommand \bibinfo  [0]{\@secondoftwo}%
\providecommand \bibfield  [0]{\@secondoftwo}%
\providecommand \translation [1]{[#1]}%
\providecommand \BibitemOpen [0]{}%
\providecommand \bibitemStop [0]{}%
\providecommand \bibitemNoStop [0]{.\EOS\space}%
\providecommand \EOS [0]{\spacefactor3000\relax}%
\providecommand \BibitemShut  [1]{\csname bibitem#1\endcsname}%
\let\auto@bib@innerbib\@empty
\bibitem [{\citenamefont {Baroni}\ \emph {et~al.}(2024{\natexlab{a}})\citenamefont {Baroni}, \citenamefont {Lamporesi},\ and\ \citenamefont {Zaccanti}}]{baroni2024}%
  \BibitemOpen
  \bibfield  {author} {\bibinfo {author} {\bibfnamefont {C.}~\bibnamefont {Baroni}}, \bibinfo {author} {\bibfnamefont {G.}~\bibnamefont {Lamporesi}},\ and\ \bibinfo {author} {\bibfnamefont {M.}~\bibnamefont {Zaccanti}},\ }\bibfield  {title} {\bibinfo {title} {Quantum mixtures of ultracold gases of neutral atoms},\ }\href {https://doi.org/10.1038/s42254-024-00773-6} {\bibfield  {journal} {\bibinfo  {journal} {Nat Rev Phys}\ }\textbf {\bibinfo {volume} {6}},\ \bibinfo {pages} {736} (\bibinfo {year} {2024}{\natexlab{a}})}\BibitemShut {NoStop}%
\bibitem [{\citenamefont {Cornish}\ \emph {et~al.}(2024)\citenamefont {Cornish}, \citenamefont {Tarbutt},\ and\ \citenamefont {Hazzard}}]{cornish2024}%
  \BibitemOpen
  \bibfield  {author} {\bibinfo {author} {\bibfnamefont {S.~L.}\ \bibnamefont {Cornish}}, \bibinfo {author} {\bibfnamefont {M.~R.}\ \bibnamefont {Tarbutt}},\ and\ \bibinfo {author} {\bibfnamefont {K.~R.~A.}\ \bibnamefont {Hazzard}},\ }\bibfield  {title} {\bibinfo {title} {Quantum computation and quantum simulation with ultracold molecules},\ }\href {https://doi.org/10.1038/s41567-024-02453-9} {\bibfield  {journal} {\bibinfo  {journal} {Nat. Phys.}\ }\textbf {\bibinfo {volume} {20}},\ \bibinfo {pages} {730} (\bibinfo {year} {2024})}\BibitemShut {NoStop}%
\bibitem [{\citenamefont {Langen}\ \emph {et~al.}(2024)\citenamefont {Langen}, \citenamefont {Valtolina}, \citenamefont {Wang},\ and\ \citenamefont {Ye}}]{langen2024}%
  \BibitemOpen
  \bibfield  {author} {\bibinfo {author} {\bibfnamefont {T.}~\bibnamefont {Langen}}, \bibinfo {author} {\bibfnamefont {G.}~\bibnamefont {Valtolina}}, \bibinfo {author} {\bibfnamefont {D.}~\bibnamefont {Wang}},\ and\ \bibinfo {author} {\bibfnamefont {J.}~\bibnamefont {Ye}},\ }\bibfield  {title} {\bibinfo {title} {Quantum state manipulation and cooling of ultracold molecules},\ }\href {https://doi.org/10.1038/s41567-024-02423-1} {\bibfield  {journal} {\bibinfo  {journal} {Nat. Phys.}\ }\textbf {\bibinfo {volume} {20}},\ \bibinfo {pages} {702} (\bibinfo {year} {2024})}\BibitemShut {NoStop}%
\bibitem [{\citenamefont {Scazza}\ \emph {et~al.}(2022)\citenamefont {Scazza}, \citenamefont {Zaccanti}, \citenamefont {Massignan}, \citenamefont {Parish},\ and\ \citenamefont {Levinsen}}]{scazza2022}%
  \BibitemOpen
  \bibfield  {author} {\bibinfo {author} {\bibfnamefont {F.}~\bibnamefont {Scazza}}, \bibinfo {author} {\bibfnamefont {M.}~\bibnamefont {Zaccanti}}, \bibinfo {author} {\bibfnamefont {P.}~\bibnamefont {Massignan}}, \bibinfo {author} {\bibfnamefont {M.~M.}\ \bibnamefont {Parish}},\ and\ \bibinfo {author} {\bibfnamefont {J.}~\bibnamefont {Levinsen}},\ }\bibfield  {title} {\bibinfo {title} {Repulsive {{Fermi}} and {{Bose Polarons}} in {{Quantum Gases}}},\ }\href {https://doi.org/10.3390/atoms10020055} {\bibfield  {journal} {\bibinfo  {journal} {Atoms}\ }\textbf {\bibinfo {volume} {10}},\ \bibinfo {pages} {55} (\bibinfo {year} {2022})}\BibitemShut {NoStop}%
\bibitem [{\citenamefont {Grusdt}\ \emph {et~al.}(2025)\citenamefont {Grusdt}, \citenamefont {Mostaan}, \citenamefont {Demler},\ and\ \citenamefont {Peña~Ardila}}]{grusdt_25}%
  \BibitemOpen
  \bibfield  {author} {\bibinfo {author} {\bibfnamefont {F.}~\bibnamefont {Grusdt}}, \bibinfo {author} {\bibfnamefont {N.}~\bibnamefont {Mostaan}}, \bibinfo {author} {\bibfnamefont {E.}~\bibnamefont {Demler}},\ and\ \bibinfo {author} {\bibfnamefont {L.~A.}\ \bibnamefont {Peña~Ardila}},\ }\bibfield  {title} {\bibinfo {title} {Impurities and polarons in bosonic quantum gases: a review on recent progress},\ }\bibfield  {journal} {\bibinfo  {journal} {Rep. Prog. Phys.}\ }\href {https://doi.org/10.1088/1361-6633/add94b} {10.1088/1361-6633/add94b} (\bibinfo {year} {2025})\BibitemShut {NoStop}%
\bibitem [{\citenamefont {Naidon}\ and\ \citenamefont {Endo}(2017)}]{naidon2017}%
  \BibitemOpen
  \bibfield  {author} {\bibinfo {author} {\bibfnamefont {P.}~\bibnamefont {Naidon}}\ and\ \bibinfo {author} {\bibfnamefont {S.}~\bibnamefont {Endo}},\ }\bibfield  {title} {\bibinfo {title} {Efimov physics: A review},\ }\href {https://doi.org/10.1088/1361-6633/aa50e8} {\bibfield  {journal} {\bibinfo  {journal} {Rep. Prog. Phys.}\ }\textbf {\bibinfo {volume} {80}},\ \bibinfo {pages} {056001} (\bibinfo {year} {2017})}\BibitemShut {NoStop}%
\bibitem [{\citenamefont {DeSalvo}\ \emph {et~al.}(2019)\citenamefont {DeSalvo}, \citenamefont {Patel}, \citenamefont {Cai},\ and\ \citenamefont {Chin}}]{desalvo2019}%
  \BibitemOpen
  \bibfield  {author} {\bibinfo {author} {\bibfnamefont {B.~J.}\ \bibnamefont {DeSalvo}}, \bibinfo {author} {\bibfnamefont {K.}~\bibnamefont {Patel}}, \bibinfo {author} {\bibfnamefont {G.}~\bibnamefont {Cai}},\ and\ \bibinfo {author} {\bibfnamefont {C.}~\bibnamefont {Chin}},\ }\bibfield  {title} {\bibinfo {title} {Observation of fermion-mediated interactions between bosonic atoms},\ }\href {https://doi.org/10/gfxx8w} {\bibfield  {journal} {\bibinfo  {journal} {Nature}\ }\textbf {\bibinfo {volume} {568}},\ \bibinfo {pages} {61} (\bibinfo {year} {2019})}\BibitemShut {NoStop}%
\bibitem [{\citenamefont {Baroni}\ \emph {et~al.}(2024{\natexlab{b}})\citenamefont {Baroni}, \citenamefont {Huang}, \citenamefont {Fritsche}, \citenamefont {Dobler}, \citenamefont {Anich}, \citenamefont {Kirilov}, \citenamefont {Grimm}, \citenamefont {{Bastarrachea-Magnani}}, \citenamefont {Massignan},\ and\ \citenamefont {Bruun}}]{baroni2024a}%
  \BibitemOpen
  \bibfield  {author} {\bibinfo {author} {\bibfnamefont {C.}~\bibnamefont {Baroni}}, \bibinfo {author} {\bibfnamefont {B.}~\bibnamefont {Huang}}, \bibinfo {author} {\bibfnamefont {I.}~\bibnamefont {Fritsche}}, \bibinfo {author} {\bibfnamefont {E.}~\bibnamefont {Dobler}}, \bibinfo {author} {\bibfnamefont {G.}~\bibnamefont {Anich}}, \bibinfo {author} {\bibfnamefont {E.}~\bibnamefont {Kirilov}}, \bibinfo {author} {\bibfnamefont {R.}~\bibnamefont {Grimm}}, \bibinfo {author} {\bibfnamefont {M.~A.}\ \bibnamefont {{Bastarrachea-Magnani}}}, \bibinfo {author} {\bibfnamefont {P.}~\bibnamefont {Massignan}},\ and\ \bibinfo {author} {\bibfnamefont {G.~M.}\ \bibnamefont {Bruun}},\ }\bibfield  {title} {\bibinfo {title} {Mediated interactions between {{Fermi}} polarons and the role of impurity quantum statistics},\ }\href {https://doi.org/10.1038/s41567-023-02248-4} {\bibfield  {journal} {\bibinfo  {journal} {Nat. Phys.}\ }\textbf {\bibinfo {volume} {20}},\ \bibinfo {pages} {68} (\bibinfo {year} {2024}{\natexlab{b}})}\BibitemShut {NoStop}%
\bibitem [{\citenamefont {Cai}\ \emph {et~al.}(2025)\citenamefont {Cai}, \citenamefont {Ando}, \citenamefont {McCusker},\ and\ \citenamefont {Chin}}]{cai_2025}%
  \BibitemOpen
  \bibfield  {author} {\bibinfo {author} {\bibfnamefont {G.}~\bibnamefont {Cai}}, \bibinfo {author} {\bibfnamefont {H.}~\bibnamefont {Ando}}, \bibinfo {author} {\bibfnamefont {S.}~\bibnamefont {McCusker}},\ and\ \bibinfo {author} {\bibfnamefont {C.}~\bibnamefont {Chin}},\ }\bibfield  {title} {\bibinfo {title} {{Fermion mediated pairing in the Ruderman-Kittel-Kasuya-Yosida to Efimov transition regime}},\ }\href {https://arxiv.org/abs/2502.06266} {\bibfield  {journal} {\bibinfo  {journal} {arXiv:2502.06266}\ } (\bibinfo {year} {2025})}\BibitemShut {NoStop}%
\bibitem [{\citenamefont {LeBlanc}\ and\ \citenamefont {Thywissen}(2007)}]{leblanc07}%
  \BibitemOpen
  \bibfield  {author} {\bibinfo {author} {\bibfnamefont {L.~J.}\ \bibnamefont {LeBlanc}}\ and\ \bibinfo {author} {\bibfnamefont {J.~H.}\ \bibnamefont {Thywissen}},\ }\bibfield  {title} {\bibinfo {title} {Species-specific optical lattices},\ }\href {https://doi.org/10.1103/physreva.75.053612} {\bibfield  {journal} {\bibinfo  {journal} {Phys. Rev. A}\ }\textbf {\bibinfo {volume} {75}},\ \bibinfo {pages} {053612} (\bibinfo {year} {2007})}\BibitemShut {NoStop}%
\bibitem [{\citenamefont {Catani}\ \emph {et~al.}(2009)\citenamefont {Catani}, \citenamefont {Barontini}, \citenamefont {Lamporesi}, \citenamefont {Rabatti}, \citenamefont {Thalhammer}, \citenamefont {Minardi}, \citenamefont {Stringari},\ and\ \citenamefont {Inguscio}}]{catani09}%
  \BibitemOpen
  \bibfield  {author} {\bibinfo {author} {\bibfnamefont {J.}~\bibnamefont {Catani}}, \bibinfo {author} {\bibfnamefont {G.}~\bibnamefont {Barontini}}, \bibinfo {author} {\bibfnamefont {G.}~\bibnamefont {Lamporesi}}, \bibinfo {author} {\bibfnamefont {F.}~\bibnamefont {Rabatti}}, \bibinfo {author} {\bibfnamefont {G.}~\bibnamefont {Thalhammer}}, \bibinfo {author} {\bibfnamefont {F.}~\bibnamefont {Minardi}}, \bibinfo {author} {\bibfnamefont {S.}~\bibnamefont {Stringari}},\ and\ \bibinfo {author} {\bibfnamefont {M.}~\bibnamefont {Inguscio}},\ }\bibfield  {title} {\bibinfo {title} {{Entropy Exchange in a Mixture of Ultracold Atoms}},\ }\href {https://doi.org/10.1103/physrevlett.103.140401} {\bibfield  {journal} {\bibinfo  {journal} {Phys. Rev. Lett.}\ }\textbf {\bibinfo {volume} {103}},\ \bibinfo {pages} {140401} (\bibinfo {year} {2009})}\BibitemShut {NoStop}%
\bibitem [{\citenamefont {Copenhaver}\ \emph {et~al.}(2019)\citenamefont {Copenhaver}, \citenamefont {Cassella}, \citenamefont {Berghaus},\ and\ \citenamefont {M\"uller}}]{copenhaver19}%
  \BibitemOpen
  \bibfield  {author} {\bibinfo {author} {\bibfnamefont {E.}~\bibnamefont {Copenhaver}}, \bibinfo {author} {\bibfnamefont {K.}~\bibnamefont {Cassella}}, \bibinfo {author} {\bibfnamefont {R.}~\bibnamefont {Berghaus}},\ and\ \bibinfo {author} {\bibfnamefont {H.}~\bibnamefont {M\"uller}},\ }\bibfield  {title} {\bibinfo {title} {{Measurement of a $^{7}\mathrm{Li}$ tune-out wavelength by phase-patterned atom interferometry}},\ }\href {https://doi.org/10.1103/PhysRevA.100.063603} {\bibfield  {journal} {\bibinfo  {journal} {Phys. Rev. A}\ }\textbf {\bibinfo {volume} {100}},\ \bibinfo {pages} {063603} (\bibinfo {year} {2019})}\BibitemShut {NoStop}%
\bibitem [{\citenamefont {Décamps}\ \emph {et~al.}(2020)\citenamefont {Décamps}, \citenamefont {Vigué}, \citenamefont {Gauguet},\ and\ \citenamefont {Büchner}}]{decamps20}%
  \BibitemOpen
  \bibfield  {author} {\bibinfo {author} {\bibfnamefont {B.}~\bibnamefont {Décamps}}, \bibinfo {author} {\bibfnamefont {J.}~\bibnamefont {Vigué}}, \bibinfo {author} {\bibfnamefont {A.}~\bibnamefont {Gauguet}},\ and\ \bibinfo {author} {\bibfnamefont {M.}~\bibnamefont {Büchner}},\ }\bibfield  {title} {\bibinfo {title} {{Measurement of the 671-nm tune-out wavelength of $^7$Li by atom interferometry}},\ }\href {https://doi.org/10.1103/physreva.101.033614} {\bibfield  {journal} {\bibinfo  {journal} {Phys. Rev. A}\ }\textbf {\bibinfo {volume} {101}},\ \bibinfo {pages} {033614} (\bibinfo {year} {2020})}\BibitemShut {NoStop}%
\bibitem [{\citenamefont {Holmgren}\ \emph {et~al.}(2012)\citenamefont {Holmgren}, \citenamefont {Trubko}, \citenamefont {Hromada},\ and\ \citenamefont {Cronin}}]{holmgren12}%
  \BibitemOpen
  \bibfield  {author} {\bibinfo {author} {\bibfnamefont {W.~F.}\ \bibnamefont {Holmgren}}, \bibinfo {author} {\bibfnamefont {R.}~\bibnamefont {Trubko}}, \bibinfo {author} {\bibfnamefont {I.}~\bibnamefont {Hromada}},\ and\ \bibinfo {author} {\bibfnamefont {A.~D.}\ \bibnamefont {Cronin}},\ }\bibfield  {title} {\bibinfo {title} {{Measurement of a Wavelength of Light for Which the Energy Shift for an Atom Vanishes}},\ }\href {https://doi.org/10.1103/PhysRevLett.109.243004} {\bibfield  {journal} {\bibinfo  {journal} {Phys. Rev. Lett.}\ }\textbf {\bibinfo {volume} {109}},\ \bibinfo {pages} {243004} (\bibinfo {year} {2012})}\BibitemShut {NoStop}%
\bibitem [{\citenamefont {McKay}\ \emph {et~al.}(2013)\citenamefont {McKay}, \citenamefont {Meldgin}, \citenamefont {Chen},\ and\ \citenamefont {DeMarco}}]{mckay2013}%
  \BibitemOpen
  \bibfield  {author} {\bibinfo {author} {\bibfnamefont {D.~C.}\ \bibnamefont {McKay}}, \bibinfo {author} {\bibfnamefont {C.}~\bibnamefont {Meldgin}}, \bibinfo {author} {\bibfnamefont {D.}~\bibnamefont {Chen}},\ and\ \bibinfo {author} {\bibfnamefont {B.}~\bibnamefont {DeMarco}},\ }\bibfield  {title} {\bibinfo {title} {{Slow Thermalization between a Lattice and Free Bose Gas}},\ }\href {https://doi.org/10/f5dxm2} {\bibfield  {journal} {\bibinfo  {journal} {Phys. Rev. Lett.}\ }\textbf {\bibinfo {volume} {111}},\ \bibinfo {pages} {063002} (\bibinfo {year} {2013})}\BibitemShut {NoStop}%
\bibitem [{\citenamefont {Trubko}\ \emph {et~al.}(2017)\citenamefont {Trubko}, \citenamefont {Gregoire}, \citenamefont {Holmgren},\ and\ \citenamefont {Cronin}}]{trubko17}%
  \BibitemOpen
  \bibfield  {author} {\bibinfo {author} {\bibfnamefont {R.}~\bibnamefont {Trubko}}, \bibinfo {author} {\bibfnamefont {M.~D.}\ \bibnamefont {Gregoire}}, \bibinfo {author} {\bibfnamefont {W.~F.}\ \bibnamefont {Holmgren}},\ and\ \bibinfo {author} {\bibfnamefont {A.~D.}\ \bibnamefont {Cronin}},\ }\bibfield  {title} {\bibinfo {title} {{Potassium tune-out-wavelength measurement using atom interferometry and a multipass optical cavity}},\ }\href {https://doi.org/10.1103/physreva.95.052507} {\bibfield  {journal} {\bibinfo  {journal} {Phys. Rev. A}\ }\textbf {\bibinfo {volume} {95}},\ \bibinfo {pages} {052507} (\bibinfo {year} {2017})}\BibitemShut {NoStop}%
\bibitem [{\citenamefont {Arora}\ \emph {et~al.}(2011)\citenamefont {Arora}, \citenamefont {Safronova},\ and\ \citenamefont {Clark}}]{arora11}%
  \BibitemOpen
  \bibfield  {author} {\bibinfo {author} {\bibfnamefont {B.}~\bibnamefont {Arora}}, \bibinfo {author} {\bibfnamefont {M.~S.}\ \bibnamefont {Safronova}},\ and\ \bibinfo {author} {\bibfnamefont {C.~W.}\ \bibnamefont {Clark}},\ }\bibfield  {title} {\bibinfo {title} {{Tune-out wavelengths of alkali-metal atoms and their applications}},\ }\href {https://doi.org/10.1103/physreva.84.043401} {\bibfield  {journal} {\bibinfo  {journal} {Phys. Rev. A}\ }\textbf {\bibinfo {volume} {84}},\ \bibinfo {pages} {043401} (\bibinfo {year} {2011})}\BibitemShut {NoStop}%
\bibitem [{\citenamefont {Herold}\ \emph {et~al.}(2012)\citenamefont {Herold}, \citenamefont {Vaidya}, \citenamefont {Li}, \citenamefont {Rolston}, \citenamefont {Porto},\ and\ \citenamefont {Safronova}}]{herold12}%
  \BibitemOpen
  \bibfield  {author} {\bibinfo {author} {\bibfnamefont {C.~D.}\ \bibnamefont {Herold}}, \bibinfo {author} {\bibfnamefont {V.~D.}\ \bibnamefont {Vaidya}}, \bibinfo {author} {\bibfnamefont {X.}~\bibnamefont {Li}}, \bibinfo {author} {\bibfnamefont {S.~L.}\ \bibnamefont {Rolston}}, \bibinfo {author} {\bibfnamefont {J.~V.}\ \bibnamefont {Porto}},\ and\ \bibinfo {author} {\bibfnamefont {M.~S.}\ \bibnamefont {Safronova}},\ }\bibfield  {title} {\bibinfo {title} {{Precision Measurement of Transition Matrix Elements via Light Shift Cancellation}},\ }\href {https://doi.org/10.1103/physrevlett.109.243003} {\bibfield  {journal} {\bibinfo  {journal} {Phys. Rev. Lett.}\ }\textbf {\bibinfo {volume} {109}},\ \bibinfo {pages} {243003} (\bibinfo {year} {2012})}\BibitemShut {NoStop}%
\bibitem [{\citenamefont {Leonard}\ \emph {et~al.}(2015)\citenamefont {Leonard}, \citenamefont {Fallon}, \citenamefont {Sackett},\ and\ \citenamefont {Safronova}}]{leonard15}%
  \BibitemOpen
  \bibfield  {author} {\bibinfo {author} {\bibfnamefont {R.~H.}\ \bibnamefont {Leonard}}, \bibinfo {author} {\bibfnamefont {A.~J.}\ \bibnamefont {Fallon}}, \bibinfo {author} {\bibfnamefont {C.~A.}\ \bibnamefont {Sackett}},\ and\ \bibinfo {author} {\bibfnamefont {M.~S.}\ \bibnamefont {Safronova}},\ }\bibfield  {title} {\bibinfo {title} {{High-precision measurements of the $^{87}\mathrm{Rb}$ $\mathrm{D}$-line tune-out wavelength}},\ }\href {https://doi.org/10.1103/PhysRevA.92.052501} {\bibfield  {journal} {\bibinfo  {journal} {Phys. Rev. A}\ }\textbf {\bibinfo {volume} {92}},\ \bibinfo {pages} {052501} (\bibinfo {year} {2015})}\BibitemShut {NoStop}%
\bibitem [{\citenamefont {Schmidt}\ \emph {et~al.}(2016)\citenamefont {Schmidt}, \citenamefont {Mayer}, \citenamefont {Hohmann}, \citenamefont {Lausch}, \citenamefont {Kindermann},\ and\ \citenamefont {Widera}}]{schmidt16}%
  \BibitemOpen
  \bibfield  {author} {\bibinfo {author} {\bibfnamefont {F.}~\bibnamefont {Schmidt}}, \bibinfo {author} {\bibfnamefont {D.}~\bibnamefont {Mayer}}, \bibinfo {author} {\bibfnamefont {M.}~\bibnamefont {Hohmann}}, \bibinfo {author} {\bibfnamefont {T.}~\bibnamefont {Lausch}}, \bibinfo {author} {\bibfnamefont {F.}~\bibnamefont {Kindermann}},\ and\ \bibinfo {author} {\bibfnamefont {A.}~\bibnamefont {Widera}},\ }\bibfield  {title} {\bibinfo {title} {{Precision measurement of the $^{87}\text{Rb}$ tune-out wavelength in the hyperfine ground state $F=1$ at 790 nm}},\ }\href {https://doi.org/10.1103/PhysRevA.93.022507} {\bibfield  {journal} {\bibinfo  {journal} {Phys. Rev. A}\ }\textbf {\bibinfo {volume} {93}},\ \bibinfo {pages} {022507} (\bibinfo {year} {2016})}\BibitemShut {NoStop}%
\bibitem [{\citenamefont {Rubio-Abadal}\ \emph {et~al.}(2019)\citenamefont {Rubio-Abadal}, \citenamefont {Choi}, \citenamefont {Zeiher}, \citenamefont {Hollerith}, \citenamefont {Rui}, \citenamefont {Bloch},\ and\ \citenamefont {Gross}}]{rubioabadal19}%
  \BibitemOpen
  \bibfield  {author} {\bibinfo {author} {\bibfnamefont {A.}~\bibnamefont {Rubio-Abadal}}, \bibinfo {author} {\bibfnamefont {J.-y.}\ \bibnamefont {Choi}}, \bibinfo {author} {\bibfnamefont {J.}~\bibnamefont {Zeiher}}, \bibinfo {author} {\bibfnamefont {S.}~\bibnamefont {Hollerith}}, \bibinfo {author} {\bibfnamefont {J.}~\bibnamefont {Rui}}, \bibinfo {author} {\bibfnamefont {I.}~\bibnamefont {Bloch}},\ and\ \bibinfo {author} {\bibfnamefont {C.}~\bibnamefont {Gross}},\ }\bibfield  {title} {\bibinfo {title} {{Many-Body Delocalization in the Presence of a Quantum Bath}},\ }\href {https://doi.org/10.1103/physrevx.9.041014} {\bibfield  {journal} {\bibinfo  {journal} {Phys. Rev. X}\ }\textbf {\bibinfo {volume} {9}},\ \bibinfo {pages} {041014} (\bibinfo {year} {2019})}\BibitemShut {NoStop}%
\bibitem [{\citenamefont {Ratkata}\ \emph {et~al.}(2021)\citenamefont {Ratkata}, \citenamefont {Gregory}, \citenamefont {Innes}, \citenamefont {Matthies}, \citenamefont {McArd}, \citenamefont {Mortlock}, \citenamefont {Safronova}, \citenamefont {Bromley},\ and\ \citenamefont {Cornish}}]{ratkata21}%
  \BibitemOpen
  \bibfield  {author} {\bibinfo {author} {\bibfnamefont {A.}~\bibnamefont {Ratkata}}, \bibinfo {author} {\bibfnamefont {P.~D.}\ \bibnamefont {Gregory}}, \bibinfo {author} {\bibfnamefont {A.~D.}\ \bibnamefont {Innes}}, \bibinfo {author} {\bibfnamefont {A.~J.}\ \bibnamefont {Matthies}}, \bibinfo {author} {\bibfnamefont {L.~A.}\ \bibnamefont {McArd}}, \bibinfo {author} {\bibfnamefont {J.~M.}\ \bibnamefont {Mortlock}}, \bibinfo {author} {\bibfnamefont {M.~S.}\ \bibnamefont {Safronova}}, \bibinfo {author} {\bibfnamefont {S.~L.}\ \bibnamefont {Bromley}},\ and\ \bibinfo {author} {\bibfnamefont {S.~L.}\ \bibnamefont {Cornish}},\ }\bibfield  {title} {\bibinfo {title} {{Measurement of the tune-out wavelength for $^{133}\mathrm{Cs}$ at 880nm}},\ }\href {https://doi.org/10.1103/physreva.104.052813} {\bibfield  {journal} {\bibinfo  {journal} {Phys. Rev. A}\ }\textbf {\bibinfo {volume} {104}},\ \bibinfo {pages} {052813} (\bibinfo {year} {2021})}\BibitemShut {NoStop}%
\bibitem [{\citenamefont {Henson}\ \emph {et~al.}(2015)\citenamefont {Henson}, \citenamefont {Khakimov}, \citenamefont {Dall}, \citenamefont {Baldwin}, \citenamefont {Tang},\ and\ \citenamefont {Truscott}}]{henson15}%
  \BibitemOpen
  \bibfield  {author} {\bibinfo {author} {\bibfnamefont {B.}~\bibnamefont {Henson}}, \bibinfo {author} {\bibfnamefont {R.}~\bibnamefont {Khakimov}}, \bibinfo {author} {\bibfnamefont {R.}~\bibnamefont {Dall}}, \bibinfo {author} {\bibfnamefont {K.}~\bibnamefont {Baldwin}}, \bibinfo {author} {\bibfnamefont {L.-Y.}\ \bibnamefont {Tang}},\ and\ \bibinfo {author} {\bibfnamefont {A.}~\bibnamefont {Truscott}},\ }\bibfield  {title} {\bibinfo {title} {{Precision Measurement for Metastable Helium Atoms of the 413nm Tune-Out Wavelength at Which the Atomic Polarizability Vanishes}},\ }\href {https://doi.org/10.1103/physrevlett.115.043004} {\bibfield  {journal} {\bibinfo  {journal} {Phys. Rev. Lett.}\ }\textbf {\bibinfo {volume} {115}},\ \bibinfo {pages} {043004} (\bibinfo {year} {2015})}\BibitemShut {NoStop}%
\bibitem [{\citenamefont {Heinz}\ \emph {et~al.}(2020)\citenamefont {Heinz}, \citenamefont {Park}, \citenamefont {Šantić}, \citenamefont {Trautmann}, \citenamefont {Porsev}, \citenamefont {Safronova}, \citenamefont {Bloch},\ and\ \citenamefont {Blatt}}]{heinz20a}%
  \BibitemOpen
  \bibfield  {author} {\bibinfo {author} {\bibfnamefont {A.}~\bibnamefont {Heinz}}, \bibinfo {author} {\bibfnamefont {A.}~\bibnamefont {Park}}, \bibinfo {author} {\bibfnamefont {N.}~\bibnamefont {Šantić}}, \bibinfo {author} {\bibfnamefont {J.}~\bibnamefont {Trautmann}}, \bibinfo {author} {\bibfnamefont {S.}~\bibnamefont {Porsev}}, \bibinfo {author} {\bibfnamefont {M.}~\bibnamefont {Safronova}}, \bibinfo {author} {\bibfnamefont {I.}~\bibnamefont {Bloch}},\ and\ \bibinfo {author} {\bibfnamefont {S.}~\bibnamefont {Blatt}},\ }\bibfield  {title} {\bibinfo {title} {{State-Dependent Optical Lattices for the Strontium Optical Qubit}},\ }\href {https://doi.org/10.1103/physrevlett.124.203201} {\bibfield  {journal} {\bibinfo  {journal} {Phys. Rev. Lett.}\ }\textbf {\bibinfo {volume} {124}},\ \bibinfo {pages} {203201} (\bibinfo {year} {2020})}\BibitemShut {NoStop}%
\bibitem [{\citenamefont {H\"ohn}\ \emph {et~al.}(2023)\citenamefont {H\"ohn}, \citenamefont {Staub}, \citenamefont {Brochier}, \citenamefont {Darkwah~Oppong},\ and\ \citenamefont {Aidelsburger}}]{hoehn23}%
  \BibitemOpen
  \bibfield  {author} {\bibinfo {author} {\bibfnamefont {T.~O.}\ \bibnamefont {H\"ohn}}, \bibinfo {author} {\bibfnamefont {E.}~\bibnamefont {Staub}}, \bibinfo {author} {\bibfnamefont {G.}~\bibnamefont {Brochier}}, \bibinfo {author} {\bibfnamefont {N.}~\bibnamefont {Darkwah~Oppong}},\ and\ \bibinfo {author} {\bibfnamefont {M.}~\bibnamefont {Aidelsburger}},\ }\bibfield  {title} {\bibinfo {title} {{State-dependent potentials for the $^{1} \text{S}_{0}$ and $^{3}\text{P}_{0}$ clock states of neutral ytterbium atoms}},\ }\href {https://doi.org/10.1103/PhysRevA.108.053325} {\bibfield  {journal} {\bibinfo  {journal} {Phys. Rev. A}\ }\textbf {\bibinfo {volume} {108}},\ \bibinfo {pages} {053325} (\bibinfo {year} {2023})}\BibitemShut {NoStop}%
\bibitem [{\citenamefont {Höhn}\ \emph {et~al.}(2024)\citenamefont {Höhn}, \citenamefont {Villela}, \citenamefont {Zu}, \citenamefont {Bezzo}, \citenamefont {Kroeze},\ and\ \citenamefont {Aidelsburger}}]{hoehn24}%
  \BibitemOpen
  \bibfield  {author} {\bibinfo {author} {\bibfnamefont {T.~O.}\ \bibnamefont {Höhn}}, \bibinfo {author} {\bibfnamefont {R.~A.}\ \bibnamefont {Villela}}, \bibinfo {author} {\bibfnamefont {E.}~\bibnamefont {Zu}}, \bibinfo {author} {\bibfnamefont {L.}~\bibnamefont {Bezzo}}, \bibinfo {author} {\bibfnamefont {R.~M.}\ \bibnamefont {Kroeze}},\ and\ \bibinfo {author} {\bibfnamefont {M.}~\bibnamefont {Aidelsburger}},\ }\bibfield  {title} {\bibinfo {title} {{Determining the $^3\text{P}_0$ excited-state tune-out wavelength of $^{174}\text{Yb}$ in a triple-magic lattice}},\ }\href {https://arxiv.org/abs/2412.14163} {\bibfield  {journal} {\bibinfo  {journal} {arXiv:2412.14163}\ } (\bibinfo {year} {2024})}\BibitemShut {NoStop}%
\bibitem [{\citenamefont {Kao}\ \emph {et~al.}(2017)\citenamefont {Kao}, \citenamefont {Tang}, \citenamefont {Burdick},\ and\ \citenamefont {Lev}}]{kao17}%
  \BibitemOpen
  \bibfield  {author} {\bibinfo {author} {\bibfnamefont {W.}~\bibnamefont {Kao}}, \bibinfo {author} {\bibfnamefont {Y.}~\bibnamefont {Tang}}, \bibinfo {author} {\bibfnamefont {N.~Q.}\ \bibnamefont {Burdick}},\ and\ \bibinfo {author} {\bibfnamefont {B.~L.}\ \bibnamefont {Lev}},\ }\bibfield  {title} {\bibinfo {title} {{Anisotropic dependence of tune-out wavelength near $\text{Dy}$ 741-nm transition}},\ }\href {https://doi.org/10.1364/oe.25.003411} {\bibfield  {journal} {\bibinfo  {journal} {Opt. Express}\ }\textbf {\bibinfo {volume} {25}},\ \bibinfo {pages} {3411} (\bibinfo {year} {2017})}\BibitemShut {NoStop}%
\bibitem [{\citenamefont {Vaidya}\ \emph {et~al.}(2015)\citenamefont {Vaidya}, \citenamefont {Tiamsuphat}, \citenamefont {Rolston},\ and\ \citenamefont {Porto}}]{vaidya2015}%
  \BibitemOpen
  \bibfield  {author} {\bibinfo {author} {\bibfnamefont {V.~D.}\ \bibnamefont {Vaidya}}, \bibinfo {author} {\bibfnamefont {J.}~\bibnamefont {Tiamsuphat}}, \bibinfo {author} {\bibfnamefont {S.~L.}\ \bibnamefont {Rolston}},\ and\ \bibinfo {author} {\bibfnamefont {J.~V.}\ \bibnamefont {Porto}},\ }\bibfield  {title} {\bibinfo {title} {{Degenerate $\text{B}$ose-$\text{F}$ermi Mixtures of Rubidium and Ytterbium}},\ }\href {https://doi.org/10/gddj9r} {\bibfield  {journal} {\bibinfo  {journal} {Phys. Rev. A}\ }\textbf {\bibinfo {volume} {92}},\ \bibinfo {pages} {043604} (\bibinfo {year} {2015})}\BibitemShut {NoStop}%
\bibitem [{\citenamefont {Hewitt}\ \emph {et~al.}(2024)\citenamefont {Hewitt}, \citenamefont {Bertheas}, \citenamefont {Jain}, \citenamefont {Nishida},\ and\ \citenamefont {Barontini}}]{hewitt2024}%
  \BibitemOpen
  \bibfield  {author} {\bibinfo {author} {\bibfnamefont {T.}~\bibnamefont {Hewitt}}, \bibinfo {author} {\bibfnamefont {T.}~\bibnamefont {Bertheas}}, \bibinfo {author} {\bibfnamefont {M.}~\bibnamefont {Jain}}, \bibinfo {author} {\bibfnamefont {Y.}~\bibnamefont {Nishida}},\ and\ \bibinfo {author} {\bibfnamefont {G.}~\bibnamefont {Barontini}},\ }\bibfield  {title} {\bibinfo {title} {Controlling the interactions in a cold atom quantum impurity system},\ }\href {https://doi.org/10.1088/2058-9565/ad4c91} {\bibfield  {journal} {\bibinfo  {journal} {Quantum Sci. Technol.}\ }\textbf {\bibinfo {volume} {9}},\ \bibinfo {pages} {035039} (\bibinfo {year} {2024})}\BibitemShut {NoStop}%
\bibitem [{\citenamefont {Lippi}\ \emph {et~al.}(2024)\citenamefont {Lippi}, \citenamefont {Gerken}, \citenamefont {H{\"a}fner}, \citenamefont {Repp}, \citenamefont {Pires}, \citenamefont {Rautenberg}, \citenamefont {Krom}, \citenamefont {Kuhnle}, \citenamefont {Tran}, \citenamefont {Ulmanis}, \citenamefont {Zhu}, \citenamefont {Chomaz},\ and\ \citenamefont {Weidem{\"u}ller}}]{lippi24}%
  \BibitemOpen
  \bibfield  {author} {\bibinfo {author} {\bibfnamefont {E.}~\bibnamefont {Lippi}}, \bibinfo {author} {\bibfnamefont {M.}~\bibnamefont {Gerken}}, \bibinfo {author} {\bibfnamefont {S.}~\bibnamefont {H{\"a}fner}}, \bibinfo {author} {\bibfnamefont {M.}~\bibnamefont {Repp}}, \bibinfo {author} {\bibfnamefont {R.}~\bibnamefont {Pires}}, \bibinfo {author} {\bibfnamefont {M.}~\bibnamefont {Rautenberg}}, \bibinfo {author} {\bibfnamefont {T.}~\bibnamefont {Krom}}, \bibinfo {author} {\bibfnamefont {E.~D.}\ \bibnamefont {Kuhnle}}, \bibinfo {author} {\bibfnamefont {B.}~\bibnamefont {Tran}}, \bibinfo {author} {\bibfnamefont {J.}~\bibnamefont {Ulmanis}}, \bibinfo {author} {\bibfnamefont {B.}~\bibnamefont {Zhu}}, \bibinfo {author} {\bibfnamefont {L.}~\bibnamefont {Chomaz}},\ and\ \bibinfo {author} {\bibfnamefont {M.}~\bibnamefont {Weidem{\"u}ller}},\ }\bibfield  {title} {\bibinfo {title} {{An Experimental Platform for Studying the Heteronuclear Efimov Effect with an Ultracold Mixture of $^6\mathrm{Li}$ and $^{133}\mathrm{Cs}$ Atoms}},\ }\href {https://doi.org/10.1007/s00601-024-01971-9} {\bibfield  {journal} {\bibinfo  {journal} {Few-Body Syst}\ }\textbf {\bibinfo {volume} {66}},\ \bibinfo {pages} {1} (\bibinfo {year} {2024})}\BibitemShut {NoStop}%
\bibitem [{\citenamefont {Becher}\ \emph {et~al.}(2018)\citenamefont {Becher}, \citenamefont {Baier}, \citenamefont {Aikawa}, \citenamefont {Lepers}, \citenamefont {Wyart}, \citenamefont {Dulieu},\ and\ \citenamefont {Ferlaino}}]{becher18}%
  \BibitemOpen
  \bibfield  {author} {\bibinfo {author} {\bibfnamefont {J.~H.}\ \bibnamefont {Becher}}, \bibinfo {author} {\bibfnamefont {S.}~\bibnamefont {Baier}}, \bibinfo {author} {\bibfnamefont {K.}~\bibnamefont {Aikawa}}, \bibinfo {author} {\bibfnamefont {M.}~\bibnamefont {Lepers}}, \bibinfo {author} {\bibfnamefont {J.-F.}\ \bibnamefont {Wyart}}, \bibinfo {author} {\bibfnamefont {O.}~\bibnamefont {Dulieu}},\ and\ \bibinfo {author} {\bibfnamefont {F.}~\bibnamefont {Ferlaino}},\ }\bibfield  {title} {\bibinfo {title} {{Anisotropic polarizability of erbium atoms}},\ }\href {https://doi.org/10.1103/PhysRevA.97.012509} {\bibfield  {journal} {\bibinfo  {journal} {Phys. Rev. A}\ }\textbf {\bibinfo {volume} {97}},\ \bibinfo {pages} {012509} (\bibinfo {year} {2018})}\BibitemShut {NoStop}%
\bibitem [{\citenamefont {Christianen}\ \emph {et~al.}(2022)\citenamefont {Christianen}, \citenamefont {Cirac},\ and\ \citenamefont {Schmidt}}]{christianen2022b}%
  \BibitemOpen
  \bibfield  {author} {\bibinfo {author} {\bibfnamefont {A.}~\bibnamefont {Christianen}}, \bibinfo {author} {\bibfnamefont {J.~I.}\ \bibnamefont {Cirac}},\ and\ \bibinfo {author} {\bibfnamefont {R.}~\bibnamefont {Schmidt}},\ }\bibfield  {title} {\bibinfo {title} {Chemistry of a {{Light Impurity}} in a {{Bose-Einstein Condensate}}},\ }\href {https://doi.org/10.1103/PhysRevLett.128.183401} {\bibfield  {journal} {\bibinfo  {journal} {Phys. Rev. Lett.}\ }\textbf {\bibinfo {volume} {128}},\ \bibinfo {pages} {183401} (\bibinfo {year} {2022})}\BibitemShut {NoStop}%
\bibitem [{\citenamefont {Onofrio}\ and\ \citenamefont {Presilla}(2002)}]{onofrio02}%
  \BibitemOpen
  \bibfield  {author} {\bibinfo {author} {\bibfnamefont {R.}~\bibnamefont {Onofrio}}\ and\ \bibinfo {author} {\bibfnamefont {C.}~\bibnamefont {Presilla}},\ }\bibfield  {title} {\bibinfo {title} {{Reaching Fermi Degeneracy in Two-Species Optical Dipole Traps}},\ }\href {https://doi.org/10.1103/physrevlett.89.100401} {\bibfield  {journal} {\bibinfo  {journal} {Phys. Rev. Lett.}\ }\textbf {\bibinfo {volume} {89}},\ \bibinfo {pages} {100401} (\bibinfo {year} {2002})}\BibitemShut {NoStop}%
\bibitem [{\citenamefont {Presilla}\ and\ \citenamefont {Onofrio}(2003{\natexlab{a}})}]{presilla03}%
  \BibitemOpen
  \bibfield  {author} {\bibinfo {author} {\bibfnamefont {C.}~\bibnamefont {Presilla}}\ and\ \bibinfo {author} {\bibfnamefont {R.}~\bibnamefont {Onofrio}},\ }\bibfield  {title} {\bibinfo {title} {{Cooling Dynamics of Ultracold Two-Species Fermi-Bose Mixtures}},\ }\href {https://doi.org/10.1103/physrevlett.90.030404} {\bibfield  {journal} {\bibinfo  {journal} {Phys. Rev. Lett.}\ }\textbf {\bibinfo {volume} {90}},\ \bibinfo {pages} {030404} (\bibinfo {year} {2003}{\natexlab{a}})}\BibitemShut {NoStop}%
\bibitem [{\citenamefont {Griessner}\ \emph {et~al.}(2006)\citenamefont {Griessner}, \citenamefont {Daley}, \citenamefont {Clark}, \citenamefont {Jaksch},\ and\ \citenamefont {Zoller}}]{griessner2006}%
  \BibitemOpen
  \bibfield  {author} {\bibinfo {author} {\bibfnamefont {A.}~\bibnamefont {Griessner}}, \bibinfo {author} {\bibfnamefont {A.~J.}\ \bibnamefont {Daley}}, \bibinfo {author} {\bibfnamefont {S.~R.}\ \bibnamefont {Clark}}, \bibinfo {author} {\bibfnamefont {D.}~\bibnamefont {Jaksch}},\ and\ \bibinfo {author} {\bibfnamefont {P.}~\bibnamefont {Zoller}},\ }\bibfield  {title} {\bibinfo {title} {{Dark-State Cooling of Atoms by Superfluid Immersion}},\ }\href {https://doi.org/10/cm4qkd} {\bibfield  {journal} {\bibinfo  {journal} {Phys. Rev. Lett.}\ }\textbf {\bibinfo {volume} {97}},\ \bibinfo {pages} {220403} (\bibinfo {year} {2006})}\BibitemShut {NoStop}%
\bibitem [{\citenamefont {Bruderer}\ \emph {et~al.}(2008)\citenamefont {Bruderer}, \citenamefont {Klein}, \citenamefont {Clark},\ and\ \citenamefont {Jaksch}}]{bruderer2008a}%
  \BibitemOpen
  \bibfield  {author} {\bibinfo {author} {\bibfnamefont {M.}~\bibnamefont {Bruderer}}, \bibinfo {author} {\bibfnamefont {A.}~\bibnamefont {Klein}}, \bibinfo {author} {\bibfnamefont {S.~R.}\ \bibnamefont {Clark}},\ and\ \bibinfo {author} {\bibfnamefont {D.}~\bibnamefont {Jaksch}},\ }\bibfield  {title} {\bibinfo {title} {{Transport of Strong-Coupling Polarons in Optical Lattices}},\ }\href {https://doi.org/10.1088/1367-2630/10/3/033015} {\bibfield  {journal} {\bibinfo  {journal} {New J. Phys.}\ }\textbf {\bibinfo {volume} {10}},\ \bibinfo {pages} {033015} (\bibinfo {year} {2008})}\BibitemShut {NoStop}%
\bibitem [{\citenamefont {Matthies}\ \emph {et~al.}(2024)\citenamefont {Matthies}, \citenamefont {Mortlock}, \citenamefont {McArd}, \citenamefont {Raghuram}, \citenamefont {Innes}, \citenamefont {Gregory}, \citenamefont {Bromley},\ and\ \citenamefont {Cornish}}]{matthies2024}%
  \BibitemOpen
  \bibfield  {author} {\bibinfo {author} {\bibfnamefont {A.~J.}\ \bibnamefont {Matthies}}, \bibinfo {author} {\bibfnamefont {J.~M.}\ \bibnamefont {Mortlock}}, \bibinfo {author} {\bibfnamefont {L.~A.}\ \bibnamefont {McArd}}, \bibinfo {author} {\bibfnamefont {A.~P.}\ \bibnamefont {Raghuram}}, \bibinfo {author} {\bibfnamefont {A.~D.}\ \bibnamefont {Innes}}, \bibinfo {author} {\bibfnamefont {P.~D.}\ \bibnamefont {Gregory}}, \bibinfo {author} {\bibfnamefont {S.~L.}\ \bibnamefont {Bromley}},\ and\ \bibinfo {author} {\bibfnamefont {S.~L.}\ \bibnamefont {Cornish}},\ }\bibfield  {title} {\bibinfo {title} {{Long-distance optical-conveyor-belt transport of ultracold $^{133}\mathrm{Cs}$ and $^{87}\mathrm{Rb}$ atoms}},\ }\href {https://doi.org/10.1103/PhysRevA.109.023321} {\bibfield  {journal} {\bibinfo  {journal} {Phys. Rev. A}\ }\textbf {\bibinfo {volume} {109}},\ \bibinfo {pages} {023321} (\bibinfo {year} {2024})}\BibitemShut {NoStop}%
\bibitem [{sm()}]{sm}%
  \BibitemOpen
  \href@noop {} {\bibinfo {title} {See {S}upplemental {M}aterial}}\BibitemShut {NoStop}%
\bibitem [{\citenamefont {Aikawa}\ \emph {et~al.}(2012)\citenamefont {Aikawa}, \citenamefont {Frisch}, \citenamefont {Mark}, \citenamefont {Baier}, \citenamefont {Rietzler}, \citenamefont {Grimm},\ and\ \citenamefont {Ferlaino}}]{aikawa12}%
  \BibitemOpen
  \bibfield  {author} {\bibinfo {author} {\bibfnamefont {K.}~\bibnamefont {Aikawa}}, \bibinfo {author} {\bibfnamefont {A.}~\bibnamefont {Frisch}}, \bibinfo {author} {\bibfnamefont {M.}~\bibnamefont {Mark}}, \bibinfo {author} {\bibfnamefont {S.}~\bibnamefont {Baier}}, \bibinfo {author} {\bibfnamefont {A.}~\bibnamefont {Rietzler}}, \bibinfo {author} {\bibfnamefont {R.}~\bibnamefont {Grimm}},\ and\ \bibinfo {author} {\bibfnamefont {F.}~\bibnamefont {Ferlaino}},\ }\bibfield  {title} {\bibinfo {title} {{Bose-Einstein Condensation of Erbium}},\ }\href {https://doi.org/10.1103/PhysRevLett.108.210401} {\bibfield  {journal} {\bibinfo  {journal} {Phys. Rev. Lett.}\ }\textbf {\bibinfo {volume} {108}},\ \bibinfo {pages} {210401} (\bibinfo {year} {2012})}\BibitemShut {NoStop}%
\bibitem [{\citenamefont {Phelps}\ \emph {et~al.}(2020)\citenamefont {Phelps}, \citenamefont {Hébert}, \citenamefont {Krahn}, \citenamefont {Dickerson}, \citenamefont {Öztürk}, \citenamefont {Ebadi}, \citenamefont {Su},\ and\ \citenamefont {Greiner}}]{phelps20}%
  \BibitemOpen
  \bibfield  {author} {\bibinfo {author} {\bibfnamefont {G.~A.}\ \bibnamefont {Phelps}}, \bibinfo {author} {\bibfnamefont {A.}~\bibnamefont {Hébert}}, \bibinfo {author} {\bibfnamefont {A.}~\bibnamefont {Krahn}}, \bibinfo {author} {\bibfnamefont {S.}~\bibnamefont {Dickerson}}, \bibinfo {author} {\bibfnamefont {F.}~\bibnamefont {Öztürk}}, \bibinfo {author} {\bibfnamefont {S.}~\bibnamefont {Ebadi}}, \bibinfo {author} {\bibfnamefont {L.}~\bibnamefont {Su}},\ and\ \bibinfo {author} {\bibfnamefont {M.}~\bibnamefont {Greiner}},\ }\bibfield  {title} {\bibinfo {title} {{Sub-second production of a quantum degenerate gas}},\ }\href {https://arxiv.org/abs/2007.10807v1} {\bibfield  {journal} {\bibinfo  {journal} {arXiv:2007.10807v1}\ } (\bibinfo {year} {2020})}\BibitemShut {NoStop}%
\bibitem [{\citenamefont {Lunden}\ \emph {et~al.}(2020)\citenamefont {Lunden}, \citenamefont {Du}, \citenamefont {Cantara}, \citenamefont {Barral}, \citenamefont {Jamison},\ and\ \citenamefont {Ketterle}}]{lunden20}%
  \BibitemOpen
  \bibfield  {author} {\bibinfo {author} {\bibfnamefont {W.}~\bibnamefont {Lunden}}, \bibinfo {author} {\bibfnamefont {L.}~\bibnamefont {Du}}, \bibinfo {author} {\bibfnamefont {M.}~\bibnamefont {Cantara}}, \bibinfo {author} {\bibfnamefont {P.}~\bibnamefont {Barral}}, \bibinfo {author} {\bibfnamefont {A.~O.}\ \bibnamefont {Jamison}},\ and\ \bibinfo {author} {\bibfnamefont {W.}~\bibnamefont {Ketterle}},\ }\bibfield  {title} {\bibinfo {title} {{Enhancing the capture velocity of a Dy magneto-optical trap with two-stage slowing}},\ }\href {https://doi.org/10.1103/physreva.101.063403} {\bibfield  {journal} {\bibinfo  {journal} {Phys. Rev. A}\ }\textbf {\bibinfo {volume} {101}},\ \bibinfo {pages} {063403} (\bibinfo {year} {2020})}\BibitemShut {NoStop}%
\bibitem [{\citenamefont {{Plotkin-Swing}}\ \emph {et~al.}(2020)\citenamefont {{Plotkin-Swing}}, \citenamefont {Wirth}, \citenamefont {Gochnauer}, \citenamefont {Rahman}, \citenamefont {McAlpine},\ and\ \citenamefont {Gupta}}]{plotkinswing20}%
  \BibitemOpen
  \bibfield  {author} {\bibinfo {author} {\bibfnamefont {B.}~\bibnamefont {{Plotkin-Swing}}}, \bibinfo {author} {\bibfnamefont {A.}~\bibnamefont {Wirth}}, \bibinfo {author} {\bibfnamefont {D.}~\bibnamefont {Gochnauer}}, \bibinfo {author} {\bibfnamefont {T.}~\bibnamefont {Rahman}}, \bibinfo {author} {\bibfnamefont {K.~E.}\ \bibnamefont {McAlpine}},\ and\ \bibinfo {author} {\bibfnamefont {S.}~\bibnamefont {Gupta}},\ }\bibfield  {title} {\bibinfo {title} {Crossed-beam slowing to enhance narrow-line ytterbium magneto-optic traps},\ }\href {https://doi.org/10.1063/5.0011361} {\bibfield  {journal} {\bibinfo  {journal} {Rev. Sci. Instrum.}\ }\textbf {\bibinfo {volume} {91}},\ \bibinfo {pages} {093201} (\bibinfo {year} {2020})}\BibitemShut {NoStop}%
\bibitem [{\citenamefont {Z{\"u}rn}\ \emph {et~al.}(2013)\citenamefont {Z{\"u}rn}, \citenamefont {Lompe}, \citenamefont {Wenz}, \citenamefont {Jochim}, \citenamefont {Julienne},\ and\ \citenamefont {Hutson}}]{zurn2013a}%
  \BibitemOpen
  \bibfield  {author} {\bibinfo {author} {\bibfnamefont {G.}~\bibnamefont {Z{\"u}rn}}, \bibinfo {author} {\bibfnamefont {T.}~\bibnamefont {Lompe}}, \bibinfo {author} {\bibfnamefont {A.}~\bibnamefont {Wenz}}, \bibinfo {author} {\bibfnamefont {S.}~\bibnamefont {Jochim}}, \bibinfo {author} {\bibfnamefont {P.}~\bibnamefont {Julienne}},\ and\ \bibinfo {author} {\bibfnamefont {J.}~\bibnamefont {Hutson}},\ }\bibfield  {title} {\bibinfo {title} {{Precise Characterization of {$^6\text{Li}$} Feshbach Resonances Using Trap-Sideband-Resolved {{RF}} Spectroscopy of Weakly Bound Molecules}},\ }\href {https://doi.org/10/gddj56} {\bibfield  {journal} {\bibinfo  {journal} {Phys. Rev. Lett.}\ }\textbf {\bibinfo {volume} {110}},\ \bibinfo {pages} {135301} (\bibinfo {year} {2013})}\BibitemShut {NoStop}%
\bibitem [{\citenamefont {Frisch}\ \emph {et~al.}(2014)\citenamefont {Frisch}, \citenamefont {Mark}, \citenamefont {Aikawa}, \citenamefont {Ferlaino}, \citenamefont {Bohn}, \citenamefont {Makrides}, \citenamefont {Petrov},\ and\ \citenamefont {Kotochigova}}]{frisch2014}%
  \BibitemOpen
  \bibfield  {author} {\bibinfo {author} {\bibfnamefont {A.}~\bibnamefont {Frisch}}, \bibinfo {author} {\bibfnamefont {M.}~\bibnamefont {Mark}}, \bibinfo {author} {\bibfnamefont {K.}~\bibnamefont {Aikawa}}, \bibinfo {author} {\bibfnamefont {F.}~\bibnamefont {Ferlaino}}, \bibinfo {author} {\bibfnamefont {J.~L.}\ \bibnamefont {Bohn}}, \bibinfo {author} {\bibfnamefont {C.}~\bibnamefont {Makrides}}, \bibinfo {author} {\bibfnamefont {A.}~\bibnamefont {Petrov}},\ and\ \bibinfo {author} {\bibfnamefont {S.}~\bibnamefont {Kotochigova}},\ }\bibfield  {title} {\bibinfo {title} {Quantum chaos in ultracold collisions of gas-phase erbium atoms},\ }\href {https://doi.org/10.1038/nature13137} {\bibfield  {journal} {\bibinfo  {journal} {Nature}\ }\textbf {\bibinfo {volume} {507}},\ \bibinfo {pages} {475} (\bibinfo {year} {2014})}\BibitemShut {NoStop}%
\bibitem [{\citenamefont {Sch{\"a}fer}\ \emph {et~al.}(2022)\citenamefont {Sch{\"a}fer}, \citenamefont {Mizukami},\ and\ \citenamefont {Takahashi}}]{schafer2022}%
  \BibitemOpen
  \bibfield  {author} {\bibinfo {author} {\bibfnamefont {F.}~\bibnamefont {Sch{\"a}fer}}, \bibinfo {author} {\bibfnamefont {N.}~\bibnamefont {Mizukami}},\ and\ \bibinfo {author} {\bibfnamefont {Y.}~\bibnamefont {Takahashi}},\ }\bibfield  {title} {\bibinfo {title} {{Feshbach Resonances of Large-Mass-Imbalance {{Er-Li}} Mixtures}},\ }\href {https://doi.org/10.1103/PhysRevA.105.012816} {\bibfield  {journal} {\bibinfo  {journal} {Phys. Rev. A}\ }\textbf {\bibinfo {volume} {105}},\ \bibinfo {pages} {012816} (\bibinfo {year} {2022})}\BibitemShut {NoStop}%
\bibitem [{\citenamefont {Presilla}\ and\ \citenamefont {Onofrio}(2003{\natexlab{b}})}]{presilla2003}%
  \BibitemOpen
  \bibfield  {author} {\bibinfo {author} {\bibfnamefont {C.}~\bibnamefont {Presilla}}\ and\ \bibinfo {author} {\bibfnamefont {R.}~\bibnamefont {Onofrio}},\ }\bibfield  {title} {\bibinfo {title} {{Cooling {{Dynamics}} of {{Ultracold Two-Species Fermi-Bose Mixtures}}}},\ }\href {https://doi.org/10.1103/PhysRevLett.90.030404} {\bibfield  {journal} {\bibinfo  {journal} {Phys. Rev. Lett.}\ }\textbf {\bibinfo {volume} {90}},\ \bibinfo {pages} {030404} (\bibinfo {year} {2003}{\natexlab{b}})}\BibitemShut {NoStop}%
\bibitem [{\citenamefont {Xie}\ \emph {et~al.}(2025)\citenamefont {Xie}, \citenamefont {Li}, \citenamefont {Zhou}, \citenamefont {Luo}, \citenamefont {Wang}, \citenamefont {Nie}, \citenamefont {Shen}, \citenamefont {Chen}, \citenamefont {Yao},\ and\ \citenamefont {Pan}}]{xie2025}%
  \BibitemOpen
  \bibfield  {author} {\bibinfo {author} {\bibfnamefont {K.}~\bibnamefont {Xie}}, \bibinfo {author} {\bibfnamefont {X.}~\bibnamefont {Li}}, \bibinfo {author} {\bibfnamefont {Y.-Y.}\ \bibnamefont {Zhou}}, \bibinfo {author} {\bibfnamefont {J.-H.}\ \bibnamefont {Luo}}, \bibinfo {author} {\bibfnamefont {S.}~\bibnamefont {Wang}}, \bibinfo {author} {\bibfnamefont {Y.-Z.}\ \bibnamefont {Nie}}, \bibinfo {author} {\bibfnamefont {H.-C.}\ \bibnamefont {Shen}}, \bibinfo {author} {\bibfnamefont {Y.-A.}\ \bibnamefont {Chen}}, \bibinfo {author} {\bibfnamefont {X.-C.}\ \bibnamefont {Yao}},\ and\ \bibinfo {author} {\bibfnamefont {J.-W.}\ \bibnamefont {Pan}},\ }\bibfield  {title} {\bibinfo {title} {{Feshbach spectroscopy of ultracold mixtures of {$^{6}{ \text{Li}}$} and {$^{164}{\text{Dy}}$} atoms}},\ }\href {https://arxiv.org/abs/2502.08099} {\bibfield  {journal} {\bibinfo  {journal} {arXiv:2502.08099}\ } (\bibinfo {year} {2025})}\BibitemShut {NoStop}%
\bibitem [{\citenamefont {Kalia}\ \emph {et~al.}(2025)\citenamefont {Kalia}, \citenamefont {Rivera}, \citenamefont {Emran}, \citenamefont {Hernandez}, \citenamefont {Kwon},\ and\ \citenamefont {Fletcher}}]{kalia2025}%
  \BibitemOpen
  \bibfield  {author} {\bibinfo {author} {\bibfnamefont {J.}~\bibnamefont {Kalia}}, \bibinfo {author} {\bibfnamefont {J.}~\bibnamefont {Rivera}}, \bibinfo {author} {\bibfnamefont {R.~R.}\ \bibnamefont {Emran}}, \bibinfo {author} {\bibfnamefont {W.~J.~S.}\ \bibnamefont {Hernandez}}, \bibinfo {author} {\bibfnamefont {K.}~\bibnamefont {Kwon}},\ and\ \bibinfo {author} {\bibfnamefont {R.~J.}\ \bibnamefont {Fletcher}},\ }\bibfield  {title} {\bibinfo {title} {{Creation of a Degenerate {{Bose-Bose}} Mixture of erbium and lithium atoms}},\ }\href {https://arxiv.org/abs/2506.00177} {\bibfield  {journal} {\bibinfo  {journal} {arXiv:2506.00177}\ } (\bibinfo {year} {2025})}\BibitemShut {NoStop}%
\bibitem [{\citenamefont {Barakhshan}\ \emph {et~al.}()\citenamefont {Barakhshan}, \citenamefont {Marrs}, \citenamefont {Bhosale}, \citenamefont {Arora}, \citenamefont {Eigenmann},\ and\ \citenamefont {Safronova}}]{UDportal}%
  \BibitemOpen
  \bibfield  {author} {\bibinfo {author} {\bibfnamefont {P.}~\bibnamefont {Barakhshan}}, \bibinfo {author} {\bibfnamefont {A.}~\bibnamefont {Marrs}}, \bibinfo {author} {\bibfnamefont {A.}~\bibnamefont {Bhosale}}, \bibinfo {author} {\bibfnamefont {B.}~\bibnamefont {Arora}}, \bibinfo {author} {\bibfnamefont {R.}~\bibnamefont {Eigenmann}},\ and\ \bibinfo {author} {\bibfnamefont {M.~S.}\ \bibnamefont {Safronova}},\ }\href@noop {} {}\bibinfo {howpublished} {{ Portal for High-Precision Atomic Data and Computation (version 2.0)}. University of Delaware, Newark, DE, USA. URL: {https://www.udel.edu/atom}}\BibitemShut {NoStop}%
\bibitem [{\citenamefont {Frisch}\ \emph {et~al.}(2013)\citenamefont {Frisch}, \citenamefont {Aikawa}, \citenamefont {Mark}, \citenamefont {Ferlaino}, \citenamefont {Berseneva},\ and\ \citenamefont {Kotochigova}}]{frisch14}%
  \BibitemOpen
  \bibfield  {author} {\bibinfo {author} {\bibfnamefont {A.}~\bibnamefont {Frisch}}, \bibinfo {author} {\bibfnamefont {K.}~\bibnamefont {Aikawa}}, \bibinfo {author} {\bibfnamefont {M.}~\bibnamefont {Mark}}, \bibinfo {author} {\bibfnamefont {F.}~\bibnamefont {Ferlaino}}, \bibinfo {author} {\bibfnamefont {E.}~\bibnamefont {Berseneva}},\ and\ \bibinfo {author} {\bibfnamefont {S.}~\bibnamefont {Kotochigova}},\ }\bibfield  {title} {\bibinfo {title} {Hyperfine structure of laser-cooling transitions in fermionic erbium-167},\ }\href {https://doi.org/10.1103/PhysRevA.88.032508} {\bibfield  {journal} {\bibinfo  {journal} {Phys. Rev. A}\ }\textbf {\bibinfo {volume} {88}},\ \bibinfo {pages} {032508} (\bibinfo {year} {2013})}\BibitemShut {NoStop}%
\bibitem [{\citenamefont {Hartog}\ \emph {et~al.}(2010)\citenamefont {Hartog}, \citenamefont {Chisholm},\ and\ \citenamefont {Lawler}}]{hartog10}%
  \BibitemOpen
  \bibfield  {author} {\bibinfo {author} {\bibfnamefont {E.~A.~D.}\ \bibnamefont {Hartog}}, \bibinfo {author} {\bibfnamefont {J.~P.}\ \bibnamefont {Chisholm}},\ and\ \bibinfo {author} {\bibfnamefont {J.~E.}\ \bibnamefont {Lawler}},\ }\bibfield  {title} {\bibinfo {title} {{Radiative lifetimes of neutral erbium}},\ }\href {https://doi.org/10.1088/0953-4075/43/15/155004} {\bibfield  {journal} {\bibinfo  {journal} {J. Phys. B: At. Mol. Opt. Phys.}\ }\textbf {\bibinfo {volume} {43}},\ \bibinfo {pages} {155004} (\bibinfo {year} {2010})}\BibitemShut {NoStop}%
\bibitem [{\citenamefont {Ban}\ \emph {et~al.}(2005)\citenamefont {Ban}, \citenamefont {Jacka}, \citenamefont {Hanssen}, \citenamefont {Reader},\ and\ \citenamefont {McClelland}}]{ban05}%
  \BibitemOpen
  \bibfield  {author} {\bibinfo {author} {\bibfnamefont {H.~Y.}\ \bibnamefont {Ban}}, \bibinfo {author} {\bibfnamefont {M.}~\bibnamefont {Jacka}}, \bibinfo {author} {\bibfnamefont {J.~L.}\ \bibnamefont {Hanssen}}, \bibinfo {author} {\bibfnamefont {J.}~\bibnamefont {Reader}},\ and\ \bibinfo {author} {\bibfnamefont {J.~J.}\ \bibnamefont {McClelland}},\ }\bibfield  {title} {\bibinfo {title} {{Laser cooling transitions in atomic erbium}},\ }\href {https://doi.org/10.1364/opex.13.003185} {\bibfield  {journal} {\bibinfo  {journal} {Opt. Express}\ }\textbf {\bibinfo {volume} {13}},\ \bibinfo {pages} {3185} (\bibinfo {year} {2005})}\BibitemShut {NoStop}%
\bibitem [{\citenamefont {Lepers}\ \emph {et~al.}(2014)\citenamefont {Lepers}, \citenamefont {Wyart},\ and\ \citenamefont {Dulieu}}]{lepers14}%
  \BibitemOpen
  \bibfield  {author} {\bibinfo {author} {\bibfnamefont {M.}~\bibnamefont {Lepers}}, \bibinfo {author} {\bibfnamefont {J.-F.}\ \bibnamefont {Wyart}},\ and\ \bibinfo {author} {\bibfnamefont {O.}~\bibnamefont {Dulieu}},\ }\bibfield  {title} {\bibinfo {title} {{Anisotropic optical trapping of ultracold erbium atoms}},\ }\href {https://doi.org/10.1103/PhysRevA.89.022505} {\bibfield  {journal} {\bibinfo  {journal} {Phys. Rev. A}\ }\textbf {\bibinfo {volume} {89}},\ \bibinfo {pages} {022505} (\bibinfo {year} {2014})}\BibitemShut {NoStop}%
\bibitem [{\citenamefont {Le~Kien}\ \emph {et~al.}(2013)\citenamefont {Le~Kien}, \citenamefont {Schneeweiss},\ and\ \citenamefont {Rauschenbeutel}}]{le_kien_2013}%
  \BibitemOpen
  \bibfield  {author} {\bibinfo {author} {\bibfnamefont {F.}~\bibnamefont {Le~Kien}}, \bibinfo {author} {\bibfnamefont {P.}~\bibnamefont {Schneeweiss}},\ and\ \bibinfo {author} {\bibfnamefont {A.}~\bibnamefont {Rauschenbeutel}},\ }\bibfield  {title} {\bibinfo {title} {{Dynamical polarizability of atoms in arbitrary light fields: general theory and application to cesium}},\ }\href {https://doi.org/10.1140/epjd/e2013-30729-x} {\bibfield  {journal} {\bibinfo  {journal} {Eur. Phys. J.}\ }\textbf {\bibinfo {volume} {67}},\ \bibinfo {pages} {92} (\bibinfo {year} {2013})}\BibitemShut {NoStop}%
\bibitem [{\citenamefont {Li}\ \emph {et~al.}(2017)\citenamefont {Li}, \citenamefont {Wyart}, \citenamefont {Dulieu},\ and\ \citenamefont {Lepers}}]{li17}%
  \BibitemOpen
  \bibfield  {author} {\bibinfo {author} {\bibfnamefont {H.}~\bibnamefont {Li}}, \bibinfo {author} {\bibfnamefont {J.-F.}\ \bibnamefont {Wyart}}, \bibinfo {author} {\bibfnamefont {O.}~\bibnamefont {Dulieu}},\ and\ \bibinfo {author} {\bibfnamefont {M.}~\bibnamefont {Lepers}},\ }\bibfield  {title} {\bibinfo {title} {{Anisotropic optical trapping as a manifestation of the complex electronic structure of ultracold lanthanide atoms: The example of holmium}},\ }\href {https://doi.org/10.1103/physreva.95.062508} {\bibfield  {journal} {\bibinfo  {journal} {Phys. Rev. A}\ }\textbf {\bibinfo {volume} {95}},\ \bibinfo {pages} {062508} (\bibinfo {year} {2017})}\BibitemShut {NoStop}%
\bibitem [{\citenamefont {Patscheider}\ \emph {et~al.}(2021)\citenamefont {Patscheider}, \citenamefont {Yang}, \citenamefont {Natale}, \citenamefont {Petter}, \citenamefont {Chomaz}, \citenamefont {Mark}, \citenamefont {Hovhannesyan}, \citenamefont {Lepers},\ and\ \citenamefont {Ferlaino}}]{patscheider_2021}%
  \BibitemOpen
  \bibfield  {author} {\bibinfo {author} {\bibfnamefont {A.}~\bibnamefont {Patscheider}}, \bibinfo {author} {\bibfnamefont {B.}~\bibnamefont {Yang}}, \bibinfo {author} {\bibfnamefont {G.}~\bibnamefont {Natale}}, \bibinfo {author} {\bibfnamefont {D.}~\bibnamefont {Petter}}, \bibinfo {author} {\bibfnamefont {L.}~\bibnamefont {Chomaz}}, \bibinfo {author} {\bibfnamefont {M.~J.}\ \bibnamefont {Mark}}, \bibinfo {author} {\bibfnamefont {G.}~\bibnamefont {Hovhannesyan}}, \bibinfo {author} {\bibfnamefont {M.}~\bibnamefont {Lepers}},\ and\ \bibinfo {author} {\bibfnamefont {F.}~\bibnamefont {Ferlaino}},\ }\bibfield  {title} {\bibinfo {title} {{Observation of a narrow inner-shell orbital transition in atomic erbium at 1299 nm}},\ }\href {https://doi.org/10.1103/PhysRevResearch.3.033256} {\bibfield  {journal} {\bibinfo  {journal} {Phys. Rev. Res.}\ }\textbf {\bibinfo {volume} {3}},\ \bibinfo {pages} {033256} (\bibinfo {year} {2021})}\BibitemShut {NoStop}%
\bibitem [{\citenamefont {De~Martino}\ \emph {et~al.}(2025)\citenamefont {De~Martino}, \citenamefont {Kiesel}, \citenamefont {Auch}, \citenamefont {Karpov},\ and\ \citenamefont {Gross}}]{zenodo}%
  \BibitemOpen
  \bibfield  {author} {\bibinfo {author} {\bibfnamefont {A.}~\bibnamefont {De~Martino}}, \bibinfo {author} {\bibfnamefont {F.}~\bibnamefont {Kiesel}}, \bibinfo {author} {\bibfnamefont {J.}~\bibnamefont {Auch}}, \bibinfo {author} {\bibfnamefont {K.}~\bibnamefont {Karpov}},\ and\ \bibinfo {author} {\bibfnamefont {C.}~\bibnamefont {Gross}},\ }\href {https://doi.org/https://doi.org/10.5281/zenodo.17249899} {\bibinfo {title} {Dissipationless tune-out trapping for a lanthanide-alkali quantum gas mixture [data set]}} (\bibinfo {year} {2025})\BibitemShut {NoStop}%
\bibitem [{\citenamefont {McAlexander}\ \emph {et~al.}(1996)\citenamefont {McAlexander}, \citenamefont {Abraham},\ and\ \citenamefont {Hulet}}]{McAlesander96}%
  \BibitemOpen
  \bibfield  {author} {\bibinfo {author} {\bibfnamefont {W.~I.}\ \bibnamefont {McAlexander}}, \bibinfo {author} {\bibfnamefont {E.~R.~I.}\ \bibnamefont {Abraham}},\ and\ \bibinfo {author} {\bibfnamefont {R.~G.}\ \bibnamefont {Hulet}},\ }\bibfield  {title} {\bibinfo {title} {{Radiative lifetime of the $2P$ state of lithium}},\ }\href {https://doi.org/10.1103/PhysRevA.54.R5} {\bibfield  {journal} {\bibinfo  {journal} {Phys. Rev. A}\ }\textbf {\bibinfo {volume} {54}},\ \bibinfo {pages} {R5} (\bibinfo {year} {1996})}\BibitemShut {NoStop}%
\end{thebibliography}%

\end{document}